\documentclass[preprints,article,accept,pdftex,moreauthors]{Definitions/mdpi} 
\firstpage{1} 
\pubvolume{1}
\issuenum{1}
\articlenumber{0}
\pubyear{2024}
\copyrightyear{2024}
\externaleditor{Academic Editor: Firstname Lastname\vspace{-12pt}}
\datereceived{12 October 2023 } 
\daterevised{23 April 2024 } 
\dateaccepted{2 May 2024 } 
\datepublished{ } 
\hreflink{https://doi.org/} 

\usepackage{makecell}[vh]

\Title{Cosmological Inference from within the Peculiar Local Universe}

\TitleCitation{Cosmological Inference from within the Peculiar Local Universe}

\Author{Roya Mohayaee 
 $^{1,}$*
\orcidA{}, Mohamed Rameez 
 $^{2}$\orcidB{} and Subir Sarkar $^{3}$\orcidC{}}

\AuthorNames{Roya Mohayaee, Mohamed Rameez and Subir Sarkar}

\AuthorCitation{Mohayaee, R.; 
 Rameez, M.; Sarkar, S.}

\address{%
$^{1}$ \quad 
 Institut d'Astrophysique de Paris, CNRS, UPMC, 98 bis Bld Arago, 75014 
 Paris, France
 \\
$^{2}$ \quad Tata Institute of Fundamental Research, Homi Bhabha Road, Mumbai 400005, India; mohamed.rameez@tifr.res.in 
\\
$^{3}$ \quad Rudolf Peierls Centre for Theoretical Physics, University of Oxford, Parks Road, Oxford OX1 3PU, UK; subir.sarkar@physics.ox.ac.uk} 

\corres{Correspondence: mohayaee@iap.fr}

\abstract{The existence of `peculiar' velocities due to the formation of cosmic structure marks a point of discord between the real universe and the usually assumed Friedmann--Lema\'{i}tre--Robertson--Walker metric, which accomodates only the smooth Hubble expansion on large scales. In the standard $\Lambda$CDM model framework, Type~Ia supernovae data are routinely ``corrected'' for the peculiar velocities of both the observer and the supernova host galaxies relative to the cosmic rest frame, in order to infer evidence for acceleration of the expansion rate from their Hubble diagram. However, observations indicate a strong, coherent local bulk flow that continues outward without decaying out to a redshift $z \gtrsim 0.1$, contrary to the $\Lambda$CDM expectation. By querying the halo catalogue of the Dark Sky Hubble-volume N-body simulation, we find that an observer placed in an unusual environment like our local universe should see correlations between supernovae in the JLA catalogue that are 2--8 times stronger  than seen by a typical or Copernican observer. This accounts for our finding that peculiar velocity corrections have a large impact on the value of the cosmological constant inferred from supernova data. We also demonstrate that local universe-like observers will infer a downward biased value of the clustering parameter $S_8$ from comparing the density and velocity fields. More realistic modelling of the peculiar local universe is thus essential for correctly interpreting cosmological data.}


\begin{document}

\section{Introduction}
\label{sec:introduction}

The flat $\Lambda$CDM `standard model' of cosmology, which has a dominant fraction of its energy density $\Omega_\Lambda \sim 0.7$ in the form of a cosmological constant and a fraction $\Omega_\text{m} \sim 0.3$ in matter (of which $\sim85\%$ is cold dark matter and $\sim 15\%$ baryons), is said to be a ``good approximation to reality'' \citep{Peebles:2022akh}. It 
 is nevertheless experiencing a crisis  due to the significant tension between the value of the present expansion rate $H_0\ (\equiv 100h$~km\,s$^{-1}$\,Mpc, $h \simeq 0.7$) determined using the `cosmic distance ladder' anchored in the local universe, and~the value inferred from the Cosmic Microwave Background (CMB) in the $\Lambda$CDM model framework~\cite{Freedman:2021ahq}. Another tension is between the growth parameter $S_8 = \sigma_8 (\Omega_\text{m}/0.3)^{0.5}$ determined from observations of weak gravitational lensing, and~from CMB data, where $\sigma_8$ is the variance of mass fluctuations in a top-hat sphere of radius $8h^{-1}$~Mpc~\cite{Abdalla:2022yfr}.

In the $\Lambda$CDM  model, data are interpreted using the Friedmann--Lema\'itre equations, obtained from general relativity assuming the Cosmological Principle. In~its modern form this assumes statistical isotropy and homogeneity in the distribution of matter and radiation in the `cosmic rest frame' (CRF) in which the CMB dipole, assumed to be of kinematical origin, is presumed to vanish. We have recently shown, however, that the distribution of distant matter as traced by quasars and radio galaxies is \emph{not} 
isotropic in the CRF~\citep{Secrest:2020has, Secrest:2022uvx}, thus challenging this  foundational assumption of the Friedmann--Lema\'{i}tre--Robertson--Walker metric. We have also shown that the acceleration of the Hubble expansion rate inferred from Type~Ia supernovae (SNe~Ia) is \emph{anisotropic} in the heliocentric frame, and so cannot be interpreted as due to $\Lambda$~\citep{Colin:2018ghy}. It is likely an artefact~\cite{Tsagas:2021dsl} because of our being `tilted' observers embedded in a  coherent bulk flow, which gives rise to the prominent dipole anisotropy in the CMB. Moreover, the local bulk flow extends out significantly further than is expected in the standard $\Lambda$CDM model, and~no convergence is seen to the CRF out as far as  $\sim 200\,h^{-1}$~Mpc~\cite{Watkins:2023rll}. In~this paper, we focus on the impact of this anomaly on the $S_8$ tension and, more importantly, on~the estimation of $\Omega_\Lambda$ from SNe~Ia~data.

The relativistic viewpoint~\citep{McClure:2007vv} is that such  deviations from the Hubble flow should be thought of as variations in the expansion velocity field of the universe, rather than as `peculiar velocities' with respect to a uniformly expanding space of the `background cosmology', which is assumed {a priori} 
 to be described by the Friedmann--Lema\'itre equations. However, it is the latter approach that has become standard in cosmology, in~particular for measuring the Hubble expansion rate. SNe~Ia data \cite{Leibundgut:2000xw}, for example, are analysed in the framework of concordance cosmology (e.g., Ref.\citep{Brout:2022vxf}) by making special relativistic corrections based on models of the local peculiar velocity field. For example, in order to obtain the `cosmological' redshift from the measured value, corrections are made for (non-Hubble) velocities using Equation~(\ref{eq:zt}). These are typically a few hundreds of km\,s$^{-1}$, so would appear to only be relevant at low redshift $z \lesssim 0.1$. However, they can affect the analysis at higher redshift too, since evidence for accelerated expansion is a dimming of the high-redshift SNe~Ia in relation to the low-redshift SNe~\cite{Perlmutter:1998np, Riess:1998cb}. Moreover, the local velocity field is quite noisy, reflecting our rather inhomogeneous neighbourhood, and~it is not clear exactly where the separation should be made between the nearby and distant universe. Various empirical methods for accounting for peculiar velocities have thus been proposed \citep{Hudson:2004et}; the above problem has often been circumvented by simply excluding from cosmological fits all the SNe~Ia at low redshifts. However this severely deprecates the sample statistics (since about a quarter of all known SNe~Ia are in fact very local) and moreover it is somewhat arbitrary, e.g.,~cuts have been applied at both $z=0.01$ and at $z=0.025$ \citep{Conley:2011ku}. Another option is to allow for an uncorrelated, and~somewhat arbitrary, dispersion in the velocities of SNe~Ia, e.g.,~\citet{Perlmutter:1998np} took the redshift uncertainty due to peculiar velocities to be $c\sigma_z = 300$~km\,s$^{-1}$, while \citet{Riess:1998cb} used $c\sigma_z = 200$~km\,s$^{-1}$.

The alternative is to correct for the peculiar velocities \citep{Neill:2007fh}: for this purpose the IRAS PSCZ catalogue \citep{Saunders:2000af}, the~SMAC catalogue of clusters \citep{Hudson:2004et} and the 2M++ catalogue \citep{Carrick:2015xza} have all been used to infer the peculiar velocity field from the underlying density field. However, these catalogues are rather limited and can be biased; moreover, linear Newtonian perturbation theory is used, hence the extracted velocities are model-dependent. Furthermore, these analyses \emph{assume} convergence to the CRF at $\gtrsim 100h^{-1}$~Mpc, as~is expected in the framework of the standard $\Lambda$CDM model, even though this is contradicted by many independent observations \citep{Hudson:2004et,Watkins:2008hf,Lavaux:2008th,Colin:2010ds,Feindt:2013pma,2016IAUS..308..336M,Howlett:2022len,Watkins:2023rll}. It is in any case inappropriate to use standard $\Lambda$CDM to make such corrections, since the model is itself a subject of the test being carried~out.

Specifically, in  analyses of the SDSS-II/SNLS3 Joint Lightcurve Analysis (JLA) catalogue of 740 SNe~Ia \citep{Betoule:2014frx}, and~the subsequent Pantheon catalogue of 1048 SNe~Ia (which includes 279 SNe~Ia from Pan-STARRS1) \citep{Scolnic:2017caz}, the~low-redshift SNe~Ia have been retained in the cosmological fits by thus ``correcting'' the individual redshifts and magnitudes of the SNe for the local `bulk flow' inferred from density field surveys out to $z \sim 0.04$ \citep{Hudson:2004et} and $z \sim 0.067$ \citep{Carrick:2015xza}. However, as noted by~\cite{Colin:2018ghy}, in~both analyses SNe~Ia immediately outside the survey volume of the peculiar velocity field were arbitrarily assumed to be at rest with respect to the cosmic rest frame, despite the fact that the same surveys detected a bulk flow extending beyond the survey volume of $372 \pm 127$~km\,s$^{-1}$ and $159 \pm 23$~km\,s$^{-1}$, respectively. Moreover the JLA and Pantheon analyses adopted different values of 150 km~s$^{-1}$ and 250~km~s$^{-1}$, respectively, for the dispersion $c\sigma_z$ of the bulk flow~velocity.

Moreover the peculiar velocity corrections applied to both JLA and Pantheon contain significant errors and inconsistencies \citep{Colin:2018ghy, Rameez:2019wdt}. Since the covariance matrices for peculiar velocity corrections have not been provided separately, the~impact of these errors on cosmological parameter estimation is hard to  quantify. 
It should also be of concern that the the residual bulk flows of the peculiar velocity surveys align approximately with the directions of maximum hemispherical asymmetry in the sky coverage of the two~catalogues.

On the theoretical side, the~correlated fluctuations of SNe~Ia magnitudes due to peculiar velocities and the impact on cosmological parameter estimation of making such corrections have been extensively studied~\cite{Hui:2005nm,Neill:2007fh,Davis:2010jq,Huterer:2015gpa}. However, all these studies assumed that the peculiar velocity statistics are those expected around a {\it typical} (aka Copernican) observer in a $\Lambda$CDM universe. Such an observer should in fact not observe a bulk flow exceeding $\sim$$200$~km\,s$^{-1}$ beyond 100~$h^{-1}$Mpc ($z \simeq 0.033$), independently of the form of the matter power spectrum \citep{Hunt:2008wp}, so clearly this assumption is in tension with reality. In~this work, we use a cosmological N-body simulations to examine the validity of the peculiar velocity covariances proposed by Hui \& Greene \citep{Hui:2005nm} in the light of additional information about our local universe. Essentially, we account for the cosmic variance due to observer location, which in our specific case is especially rare for a $\Lambda$CDM universe---as we quantify~below.

The bulk flow observations suggest that we are \emph{not} typical observers in a $\Lambda$CDM universe \citep{Hellwing:2016pdl, Hellwing:2018tiq, Rameez:2017euv}. We discuss here the correlated fluctuations of SNe~Ia magnitudes and redshifts due to the peculiar velocities and bulk flows in and around `Local Universe (LU)-like' environments in the $z=0$ halo catalogue of the DarkSky $\Lambda$CDM simulations \citep{Skillman:2014qca}. We find that previous theoretical predictions \citep{Hui:2005nm} for randomly selected typical observers  have underestimated the actual covariances for observers like ourselves by a factor of~2--8.

Next we discuss (Section~\ref{sec:PVSN1a}) the peculiar velocity corrections employed in JLA \citep{Betoule:2014frx} and show that these are both arbitrary and incomplete (Section~\ref{sec:JLA}). We compare the magnitude of the velocities used for the corrections in JLA against those obtained from the Cosmicflows-3 (CF-3) compilation \citep{Tully:2016ppz} and demonstrate that the JLA values are underestimated by $\sim$48\% on average. We also review the various relevant sources of uncertainties and dispersions that go into the JLA cosmological fits. We then explore (Section~\ref{sec:likelihood0}) various methods to fit for the extent of the bulk flow in the LU and present our likelihood analysis (Section~\ref{sec:likelihood1}). This is followed by a discussion of related work (Section~\ref{sec:discussion}). An~Appendix presents the standard methodology of cosmology with SNe~Ia (Section~\ref{SNcosmo}), and~the JLA catalogue (Appendix~\ref{JLA}).

We find that for any consistent treatment of the peculiar velocities (including ignoring them altogether), the~JLA dataset favours $\Omega_{\Lambda} \lesssim 0.45$ and is consistent with a non-accelerating universe at $\sim$$2\sigma$. Larger values of $\Omega_{\Lambda}$ which have been found in other analyses of the JLA catalogue \citep{Nielsen:2015pga, Rubin:2016iqe} are in fact due to the incomplete peculiar velocity ``corrections'' applied. We demonstrate that, consistent with the recent finding from the Cosmicflows-4 survey~\cite{Watkins:2023rll}, the~JLA data favour a fast ($>$$250$~km\,s$^{-1}$) bulk flow extending out to $>$$200h^{-1}$~Mpc, which is quite unexpected in the standard $\Lambda$CDM~model.

Subsequently (Section~\ref{sec:S8}), we examine the $S_8$ parameter inferred by randomly selected Copernican observers as well as constrained `Local-Universe like' observers by comparing the peculiar velocities around them with those expected from the density field. While the variance in $S_8$ as inferred by Copernican observers is already larger than the disagreement between a recent measurement \cite{Boruah:2019icj} and the $\Lambda$CDM fiducial value from  Planck, LU-like observers see both an additional downward bias on $S_8$ and a larger cosmic~variance.

\section{Selecting Local Universe Like~Environments}
\label{sec:PVSN1a}
Making the usual assumption that the CMB dipole is due to our motion with respect to the CRF (also known as the `CMB frame') in which the universe looks isotropic, so the luminosity distance $d_\textrm{L}$ is related to the redshift $z$ as in Equation~(\ref{eq:DLEQ}), 
the redshift of a supernova in the heliocentric frame $z_\textrm{hel}$ (obtained by correcting the actually measured redshift for the Earth's motion around the Sun) is related to its redshift $z$ in the CMB frame (sometimes labelled $z_\mathrm{CMB}$) as~\cite{Davis:2010jq}:
\begin{equation}
    \label{eq:zt}
    1+z_\textrm{hel}  =  (1+z_{\odot})\times(1+z_\textrm{SN})\times (1+z)\,\, ,
\end{equation}
where $z_{\odot}$ is the redshift induced by our motion with respect to the CMB and $z_\textrm{SN}$ is the redshift due to the peculiar motion of supernova host galaxy in the CMB frame. The~luminosity distance is similarly corrected as:
\begin{equation}
\label{eq:dlt}
d_\textrm{L} (z_\textrm{hel}) = d_\textrm{L}(z) (1 + z_{\odot})\times(1+z_\textrm{SN})^2
\end{equation} 
to obtain $d_\textrm{L}$ as a function of $z$ Equation (\ref{eq:DLEQ}) for the standard $\Lambda$CDM model (see Appendix~\ref{SNcosmo}). The~covariance of SNe~Ia magnitudes due to peculiar velocities is then given by \citep{Hui:2005nm, Davis:2010jq, Huterer:2015gpa}:
\begin{equation}
  \label{eq:Sij}    
S_{ij} = \langle \delta m_i \delta m_j \rangle = \bigg[ \frac{5}{\rm{ln}10}\bigg]^2 
\frac{(1+z_i)^2}{H(z_i)d_\textrm{L}(z_i)}\frac{(1+z_j)^2}{H(z_j)d_\textrm{L}(z_j)} \xi_{ij},
\end{equation}
where
\begin{multline}
    \label{eq:Vij}    
\xi_{ij}=\langle (\vec{v}_i . \hat{n}_i) (\vec{v}_j. \hat{n}_j) \rangle = \frac{\textrm{d}D_i}{d \tau} 
\frac{\textrm{d}D_j}{d \tau} \int \frac{\textrm{d}k}{2\pi^2} P(k, a=1)\\ \times \sum_l (2l + 1) j'_l(k\chi_i) j'_l(k\chi_j)P_l(\hat{n}_i . \hat{n}_j).
\end{multline}
Here 
$D_i$ is the linear structure growth factor at the redshift of the $i\textrm{th}$ SNe, $ j'_l$ is the derivative of the $l\textrm{th}$ spherical Bessel function and $P_l$ is the Legendre polynomial of order $l$. Note that according to this expression, the covariance in magnitudes between two SNe depends only on their relative angular separation (which comes in through $P_l$) and is independent of their absolute~directions.

\begin{figure}[H]
 \includegraphics[width=0.49\columnwidth]{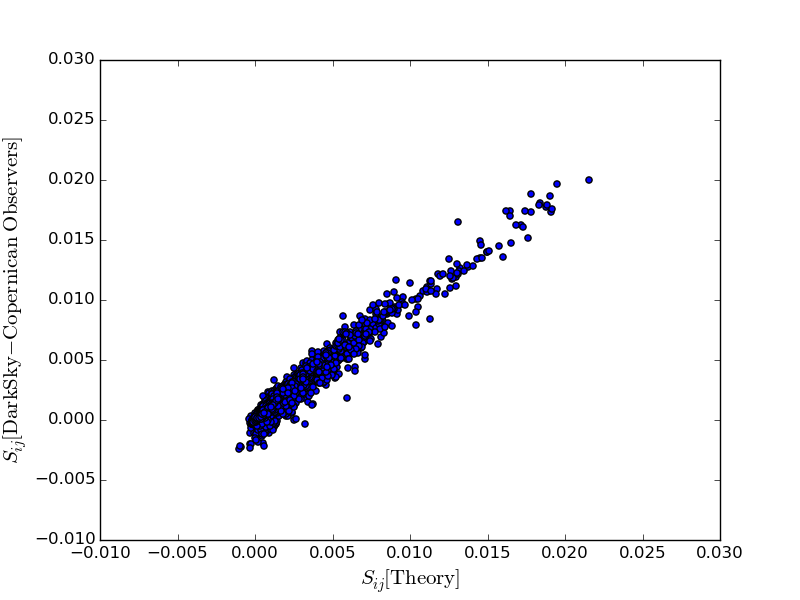}
 \includegraphics[width=0.49\columnwidth]{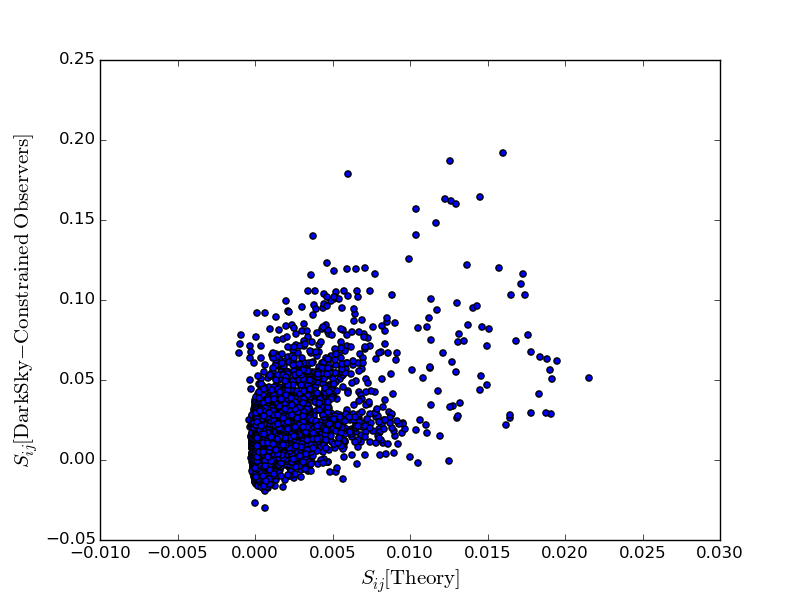}
 \caption{The theoretically expected covariance $S_{ij}$ (Equation \eqref{eq:Sij}) plotted against the value found in N-body simulations---in regions around typical observers (\textbf{left}) and constrained `Local Universe-like' observers (\textbf{right}). Each point is an average over 1000~observers.}
  \label{fig:DSvsTH}
\end{figure}

N-body simulations can be used to estimate $\xi_{ij}$ Equation~(\ref{eq:Sij}) for different assumed observers. Figure~\ref{fig:DSvsTH} compares $S_{ij}$ evaluated using the $\Lambda$CDM expectation for $\xi_{ij}$ with that read off the $z = 0$ snapshot halo catalogue of \texttt{Dark Sky 
}, a~Hubble volume, trillion-particle simulation \citep{Skillman:2014qca}, for~two very different classes of observers. For~the `Copernican observer' in Figure~\ref{fig:DSvsTH} (left), the~halo containing the observer and its orientation are selected at random---such an observer sees the universe as isotropic and homogeneous. However, for the constrained `LU-like' observer' (right), only halos satisfying the following criteria are considered (the first three being the same as in~Ref.\cite{Hellwing:2016pdl}):

\begin{itemize}
\item[(i)] The observer halo has a Milky Way (MW)-like mass, in~the range $2.2 \times 10^{11} < M_{200} < 1.4 \times 10^{12} M_\odot$ \citep{Cautun:2014dda} for the halo mass contained within 200 kpc.
\item[(ii)]  The bulk velocity in a sphere of radius $R = 3.125h^{-1}$~Mpc centred on the observer is $V= 622 \pm 150$~km\,s$^{-1}$ 
\item[(iii)]  A Virgo-cluster like halo of mass $M = (1.2 \pm 0.6) \times 10^{15} h^{-1} M_\odot$ is present at a distance $D = 12 \pm 4 h^{-1}$~Mpc from the observer.
\item[(iv)]  The angle between the bulk flow of (ii) and the direction to the Virgo-like halo of (iii) is $(44.5 \pm 5)^\circ$ 
\item[(v)]  The bulk velocity in a sphere of $R=180h^{-1}$~Mpc centered on the observer is 
$260 \pm 100$ ~km\,s $^{-1}$~\citep{Colin:2010ds}.
\item[(vi)]  The angle between the bulk flow of (v) and the direction to the Virgo-like halo of (iii) is $(69.9 \pm 7.5)^\circ$.  
\item[(vii)]  The angle between the bulk flows of (ii) and (v) is $(35.6 \pm 7.5)^\circ$.
\end{itemize}
Note 
 that such observers are very rare---we had to examine 8,721,498 halos in order to find 1000 that satisfied these criteria. Hence the above probability of 1000/8721498 = 0.0115\% is an upper limit on the probability of finding our Local Universe in $\Lambda$CDM~cosmology. After an observer satisfying the above criteria is found, the~entire system is rotated so that the direction of the bulk flow of criterion (ii) and the direction to the Virgo-like halo of criterion (iii) correspond to the real observed directions. The~criterion on the bulk flow direction is exact, while the criterion on the direction to the Virgolike halo is imposed only on the azimuthal angle in a coordinate system in which the $z$-axis points towards the bulk flow direction. Criterion (iv) then suffices to orient the system. Note that the angular tolerances in (iv), (vi), and (vii) are set to be less stringent than actual observational constraints in~order to limit the required computation time. \vspace{-12pt}

Subsequently, the~halos closest to each JLA supernova with $z < 0.1$ (of which there are 152) are identified in the DarkSky simulation as a proxy for the supernova host galaxy, and~their velocities are queried. From~these velocities, $\xi_{ij}$ can be calculated for each pair in the set of 152 supernovae. For~the typical observer of Figure~\ref{fig:DSvsTH} (left), none of the steps regarding directional orientation discussed above are considered and observers are simply picked at random. As~seen in Figure~\ref{fig:DSvsTH} (right), a~realistic LU-like observer sees on average correlations between the supernovae of a JLA-like catalogue that are 2--8 times \emph{stronger} than does a typical observer. This illustrates that the theoretical covariances of \citet{Hui:2005nm}, 
as~given in Equation~(\ref{eq:Vij}), is valid only for idealised observers who see neither a local bulk flow nor a preferred orientation in the~sky.

\subsection{Peculiar Velocity Corrections in~JLA}
\label{sec:JLA}
\begin{figure}[H]
 \includegraphics[width=0.5\columnwidth]{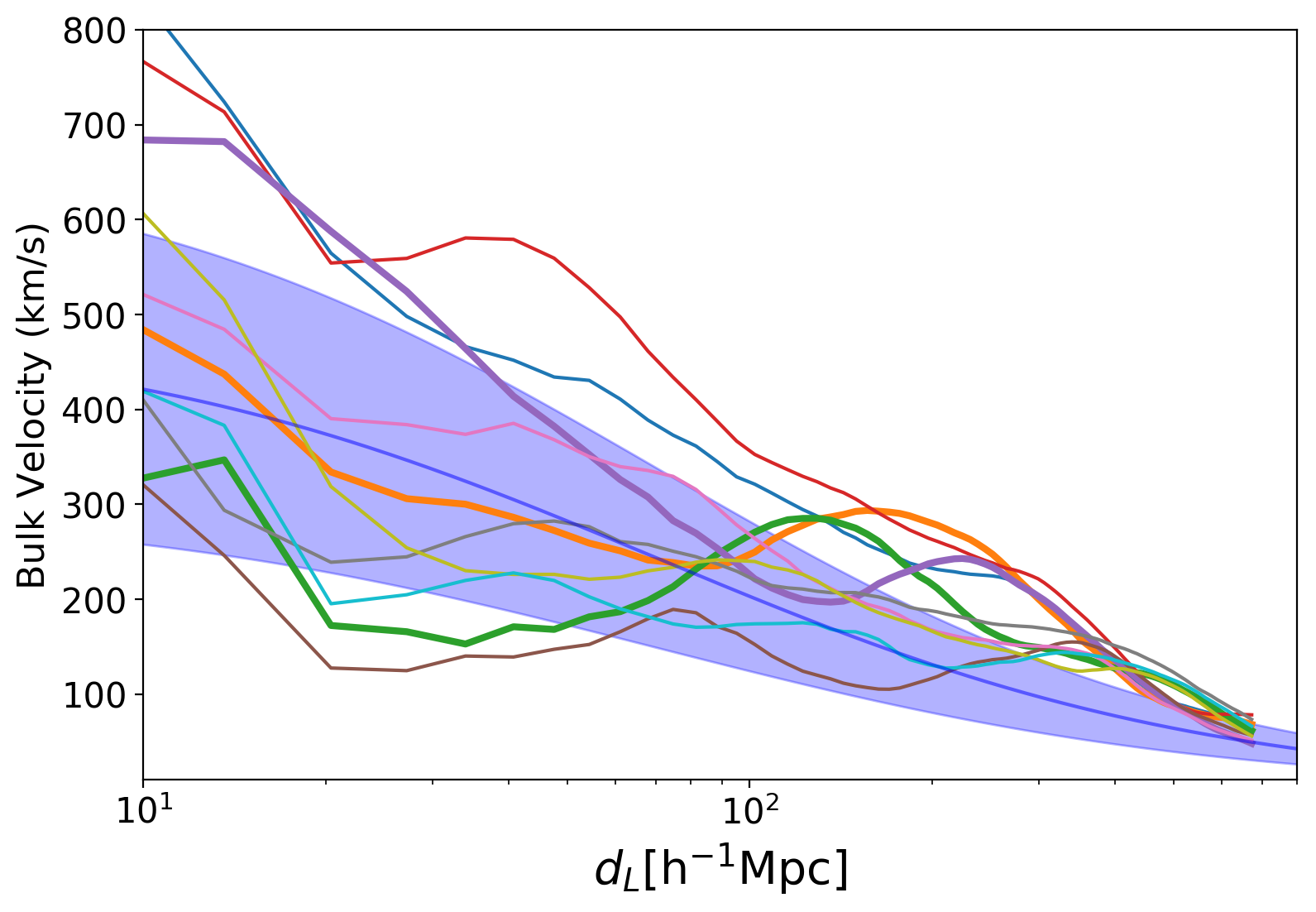}
  \caption{The 
 bulk flow velocity profiles around 10 random `Local Universe-like' observers satisfying the criteria in Section~\ref{sec:PVSN1a}. Note that the velocity profile around an individual observer need not decrease monotonically, even though the ensemble average in the $\Lambda$CDM model (dark blue curve) does so. Indeed, the Cosmicflows-4 data~\cite{Watkins:2023rll} indicates a rising velocity with depth. The~shaded blue region is the $\pm 1 \sigma$ band around the mean~value.}
 \label{fig:NonCopProfiles}
\end{figure}

It had been noted in Ref.~\cite{Colin:2018ghy} that the peculiar velocity `corrections' applied to the SNe~Ia redshifts and magnitudes in the JLA catalogue (see Appendix~\ref{JLA}) are neither consistent nor complete. SNe~Ia immediately beyond $z \sim 0.06$ were taken to be stationary with respect to the CMB and assumed to only have an uncorrelated velocity dispersion $c \sigma_z=$ 150 km s$^{-1}$ in the cosmological fits, even though observations of clusters indicate a bulk velocity of $372 \pm 127$ km~s$^{-1}$ due to sources beyond $200h^{-1}$~Mpc \citep{Hudson:2004et}. Unlike the intrinsic dispersion $\sigma_{M_0}$ which is assumed to be redshift-\emph{independent}, the~dispersion in the magnitudes as a result of the velocity dispersion is $5 \sigma_z /(z \mathrm{ln}10)$ i.e.,~the magnitudes of lower redshift supernovae are selectively more dispersed. As~seen in Figure~\ref{fig:NonCopProfiles}, the~typical bulk flow in a $\Lambda$CDM universe (see, e.g., Ref.\citep{Hong:2014jla}) continues to much larger distances, with~the velocity decreasing gradually. In~some environments, the~bulk velocity may even increase beyond a certain scale (as seen in the CosmicFlow-4 survey \citep{Watkins:2023rll}), although~the overall trend has to be a decreasing~one if the Universe is to become homogeneous when averaged on large scales.

\begin{figure}[H]
  \includegraphics[width=0.5\columnwidth]{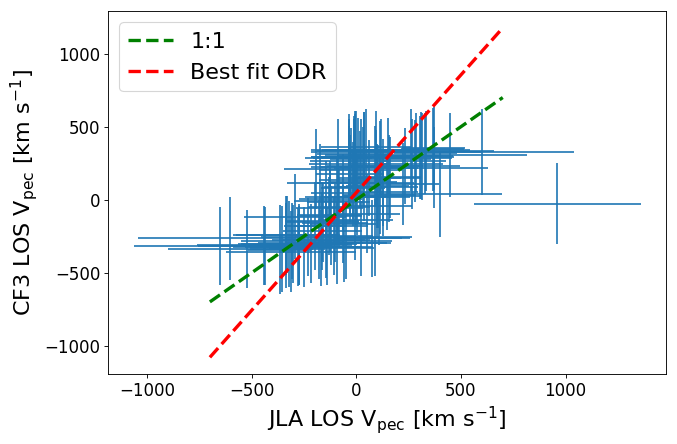}
  \caption{The 
~line-of-sight velocity of SNe~Ia inferred from their $z_\textrm{hel}$ and $z_\textrm{CMB}$ values quoted by JLA, plotted versus the line-of-sight component of the velocity of the group the object belongs to in the  dataset ($\langle{V_\mathrm{CMB}}\rangle - gp$). The~blue horizontal bars are the diagonal errors in the JLA cosmology fit (statistical plus systematic), while the blue vertical bars indicate the random error of 250 km s$^{-1}$ in the  measurement. The~green dashed line indicates when the two are equal, while the red dashed line shows the best-fit orthogonal distance regression which has a slope of 1.61, i.e.,~the JLA velocities have been underestimated on average by 48\%. (Note that the outlier (SN1992bh) has a peculiar velocity of $\sim1000$~km\,s$^{-1}$ according to JLA, but~zero according to CF-3.)}
  \label{fig:JLAvsCF}
\end{figure}

Cosmicflows-3 (CF-3) \citep{Tully:2016ppz} is a compilation of the peculiar velocities of 17,669 nearby galaxies, using various independent distance estimators such as the Tully-Fisher relationship. In~Figure~\ref{fig:JLAvsCF}, we compare the velocities that have been used to correct the JLA redshifts with those from CF-3. The~ galaxy in the CF-3 dataset corresponding to a JLA supernova is identified by cross-matching with a tolerance of $0.01^\circ$
, using the tool \texttt{k3match}. Out of 119 JLA SNe~Ia at $z_\textrm{CMB}<0.06$, 112 have CF-3 counterparts within $0.01^\circ$. It is seen from the regression line \citep{ODR} in Figure~\ref{fig:JLAvsCF} that peculiar velocities have been systematically \emph{underestimated} by 48\% in the corrections applied in the JLA analysis \citep{Betoule:2014frx}, compared to the actual measurements by~CF-3.

\unskip

\subsection{Fitting for a Bulk~Flow}
\label{sec:likelihood0}
\begin{figure}[H]
  \includegraphics[width=0.7\columnwidth]{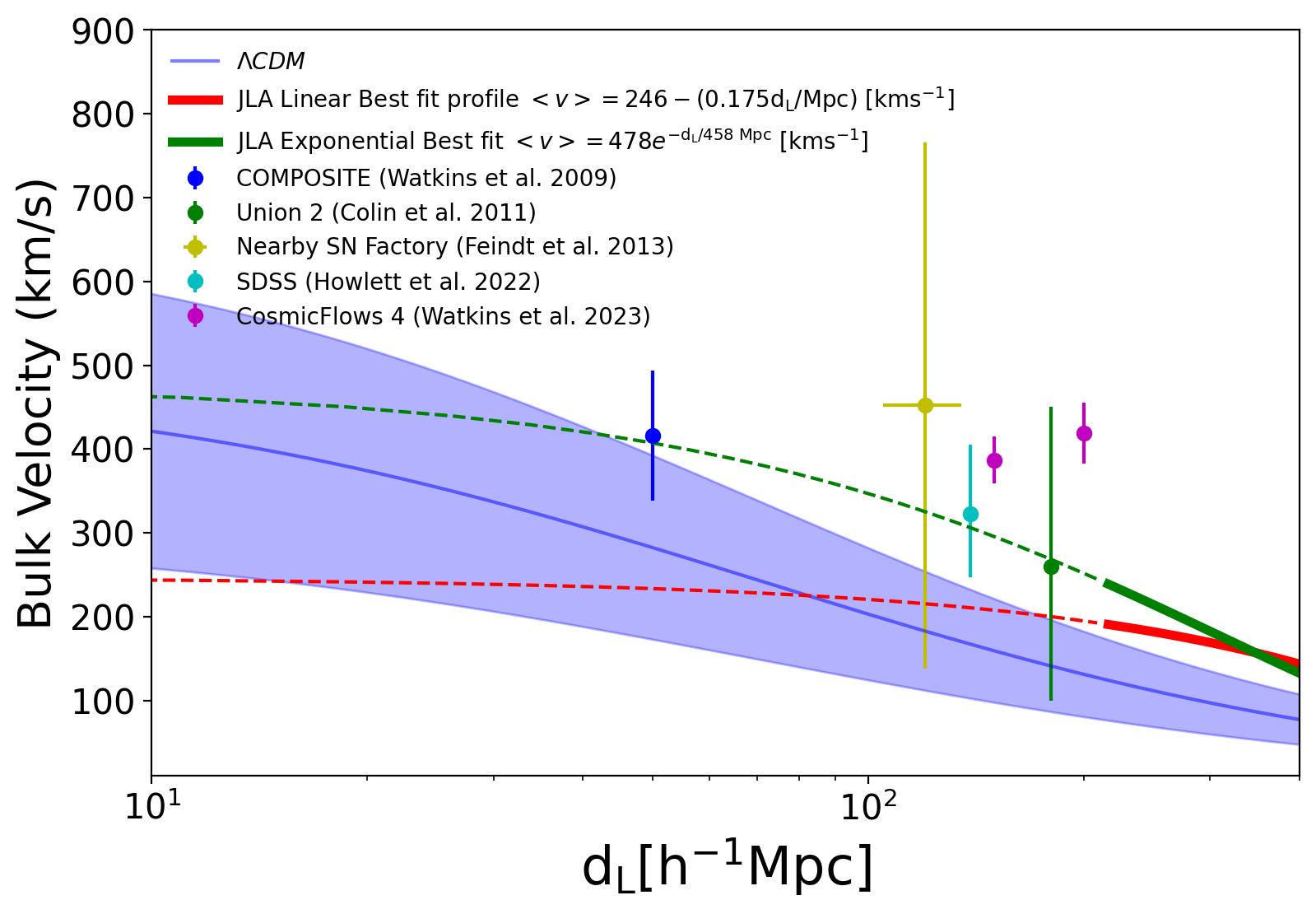}
  \caption{The 
 profile of bulk flow expected in a $\Lambda$CDM universe using
  a top-hat window function is shown as the solid blue line, while  the shaded region shows the $\pm1\sigma$ range. Several discrepant measurements (with $\pm1\sigma$ uncertainties) using various surveys are shown for comparison \cite{Watkins:2008hf, Colin:2010ds, Feindt:2013pma, Howlett:2022len, Watkins:2023rll}. The~red and green lines show, respectively, the~linear and exponential fits to the bulk flow using JLA data --- rows 9 and 10 of Table~\ref{tab:results3} (the dashed lines indicate the extrapolated fits). Note that the fits are \emph{not} done to the other selected measurements shown.}
\label{fig:lcdmlinearexponential}
\end{figure}

We consider two illustrative profiles for the bulk flow velocity. A~linear $\sim$$1/r$ fall-off is expected if we are, e.g.,~being pulled by a `Great Attractor':
\begin{equation}
\label{eq:PQlin}
  \langle v \rangle  = P - Q' d_\textrm{L} ,
\end{equation} 
where $Q'$ is a (dimensionless) scale parameter. We ensure that the velocity never goes negative by setting it to zero above $d_\textrm{L}= P/Q'$ (see Figure~\ref{fig:lcdmlinearexponential}). Our unusual bulk flow may however be due to a different physical cause, so we also consider an 
an exponentially falling form
\begin{equation}
\label{eq:PQexp}
  \langle v \rangle  = P \mathrm{e}^{-d_\textrm{L}/Q} ,
\end{equation} 
where $Q$ is the scale of the flow. The~free parameters $P$ and $Q$ or $Q'$ in our modelling of the bulk flow can be determined along with the 10 other parameters used to fit SNe~Ia~data.

The effect of a bulk flow term is to modify the distance modulus of Equation~(\ref{eq:DMOD}) such that:
\begin{equation}
\label{eq:bulkmagres}
\Delta m^\textrm{bulk}(P,Q, z_i) = - \left( \frac{5}{\textrm{ln} 10} \right) \frac{(1+z_i)^2}{H(z_i)d_\textrm{L}(z_i)} \hat{n}_i [P \textrm{e}^{-d_\textrm{L}(z_i)/Q}] ,
\end{equation} 
for the exponentially falling bulk flow, and~likewise for the linearly falling bulk flow with the expression from Equation~(\ref{eq:PQlin}) now substituted in the square~brackets.

\subsection{The Likelihood~Analysis}
\label{sec:likelihood1}
We can now rewrite  $z_\textrm{SN}$ and $z$ in Equation~(\ref{eq:zt}) as functions of $z_\textrm{hel}$, $P$ and $Q$ for the exponential (Equation~\eqref{eq:PQexp}) or $Q'$ for the linear (Equation~\eqref{eq:PQlin}) bulk flow models, respectively. A~MLE~\cite{Nielsen:2015pga} is then used with the two additional parameters $P$ and $Q$ (or $Q'$) for the bulk flow, in~addition to the usual light curve fitting parameters in the SALT2 template \citep{Betoule:2014frx}: $\alpha, x_{1,0}, \sigma_{x_{1,0}}, \beta, c_0, \sigma_{c_0}, M_0, \sigma_{M_0}$. Along with the two $\Lambda$CDM model parameters $\Omega_\textrm{m}$ and $\Omega_\Lambda$ we then have 12 parameters in total. As~shown in Table~\ref{tab:results3}, the~following fits are performed (including an additional dispersion of $c\sigma_z = 150$ km s$^{-1}$ as recommended by Ref.~\cite{Betoule:2014frx}):

\begin{enumerate}

\item The same 10-parameter fit as in Ref.~\cite{Nielsen:2015pga}, using only the $z_\textrm{CMB}$ values provided by~JLA.

\item The 10-parameter fit using $d_\textrm{L} (z_\textrm{CMB}, z_\textrm{hel})= [(1+z_\textrm{hel})/(1+z_\textrm{CMB})] d_\textrm{L}(z_\textrm{CMB})$---as used in Ref.~\cite{Davis:2010jq}---and the JLA provided $z_\textrm{hel}$ and $z_\textrm{CMB}$ values. 

\item The 10-parameter fit as in 1, using only $z_\textrm{hel}$ values (JLA provided) as was done in all SNe~Ia analyses until~2011.

\item The 10-parameter fit as in 1, using JLA provided $z_\textrm{hel}$ values, after~subtracting out bias corrections to $m^*_B$.

\item Exponentially falling bulk flow: 12-parameter fit (including the $P$ and $Q$ parameters of Equation~(\ref{eq:PQexp}), using only JLA provided $z_\textrm{hel}$ values. No peculiar velocity corrections are~applied.

\item Linearly falling bulk flow: 12-parameter fit (including the $P$ and $Q$ parameters of Equation~(\ref{eq:PQlin}) using only JLA provided $z_\textrm{hel}$ values. No peculiar velocity corrections are~applied.

\item JLA-corrected redshifts + Exponential bulk flow: 12-parameter fit: SNe~Ia with peculiar velocity corrections applied by JLA, are treated as in (ii) above, while an exponentially falling bulk flow is fitted to the remaining~SNe.

\item JLA-corrected redshifts + Linear bulk flow: As in 7, but~with the linear parametrisation of the bulk~flow.

\item CF-3 data and the exponential bulk flow fit: 12-parameter fit using Equation~(\ref{eq:dlt}) with the -derived values of $z_\textrm{hel}$ and $z_\textrm{CMB}$ (see Section~\ref{sec:JLA}) used for the low $z$ SNe~Ia to which the velocity correction can be applied. For~the remaining objects, we use the JLA $z_\textrm{hel}$ values, and~an Exponential bulk flow is fitted using Equation~(\ref{eq:PQexp}) as described~above.

\item CF-3 data and the linear bulk flow fit: 12-parameter fit using Equation~(\ref{eq:PQlin}) with the -derived values of $z_\textrm{hel}$ and $z_\textrm{CMB}$ (see Section~\ref{sec:JLA})  used for the low $z$ SNe~Ia to which the velocity correction can be applied. For~the remaining objects, we use the JLA $z_\textrm{hel}$ values, and~a linear bulk flow is fitted using Equation~(\ref{eq:PQlin}).

\end{enumerate}

In all these fits, the~direction of the bulk flow is fixed to be in the CMB dipole direction as most previous analyses have shown large dipoles at intermediate redshifts converging to this direction \citep{Watkins:2008hf,Lavaux:2008th,Colin:2010ds,Colin:2017juj,Rameez:2017euv}. In~Table~\ref{tab:results3}, we also show the fit results after imposing the additional constraint of `No acceleration' for a $\Lambda$CDM universe i.e.: $q_0 \equiv \Omega_\Lambda/2 - \Omega_\textrm{m} = 0$. For~the last two fits, we also show the effect of imposing the constraint of zero curvature (`Flat'): i.e.,~$\Omega_\Lambda + \Omega_\textrm{m} = 1$.

The bulk flow fit is $\langle v \rangle  = 478 \mathrm{e}^{-d_\textrm{L}/458~\mathrm{Mpc}} \mathrm{km~s}^{-1}$ for the exponential decay form (\ref{eq:PQexp}), and~$\langle v \rangle = [246 - 0.175 (d_\textrm{L}/\mathrm{Mpc})] \mathrm{km~s}^{-1}$ for the linearly falling form (\ref{eq:PQlin}). Including the bulk flow \emph{always} improves the quality of the fit as can be seen from the smaller values of $-2 \log {\cal L}_\textrm{max}$. This justifies adding the two parameters characterising it. In~all the above fits apart from the `No acceleration' ones, the~best-fit bulk flow extends beyond $200h^{-1}$~Mpc at $250$~km\,s$^{-1}$. Figure~\ref{fig:lcdmlinearexponential} shows our results along with selected recent~observations.

\begin{figure}[H]
\includegraphics[width=0.45\columnwidth]{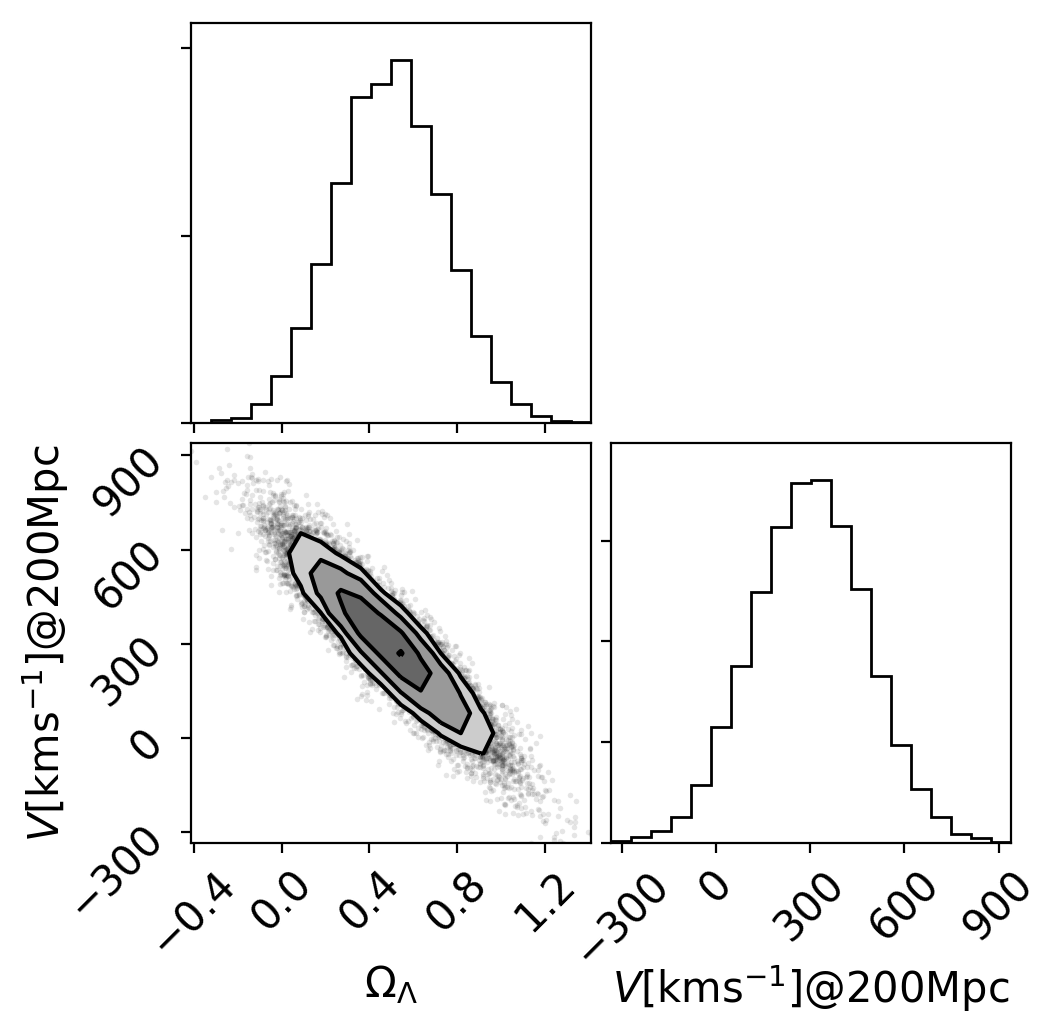}
\caption{1, 
 2 and 3 $\sigma$ contours corresponding to the fit in row 9 of Table~\ref{tab:results3} wherein peculiar velocities from Cosmicflows-3 are used for SN-by-SN corrections and the flow is allowed to continue beyond the survey volume with an exponential fall-off  (Equation~\ref{eq:PQexp}). The~velocity of the bulk flow at a top-hat smoothing scale of radius 200~Mpc is shown in the right histogram of the posterior. The top histogram shows the extracted value of $\Omega_\Lambda$, which is seen to be degenerate with this bulk flow velocity.}
\label{fig:LCDMBFposterior}
\end{figure}

Using the CF-3 data and the linear bulk flow fit, as~well as other fits of similar quality, the~difference in the goodness of fit of the best model (with the lowest value of $-2 \log {\cal L}_\textrm{max}$) with respect to the corresponding `No acceleration' fit is now significantly smaller compared to previous studies. Figure~\ref{fig:LCDMBFposterior}
demonstrates the degeneracy between the derived value of $\Omega_\Lambda$ and the local bulk flow, illustrating that the latter is an essential nuisance parameter to be added to cosmological fits when analysing SNe~Ia. Allowing for the bulk flow in the fit demonstrates that the evidence for acceleration using SNe~Ia data alone is even weaker than was found previously \citep{Nielsen:2015pga}.

\begin{table}[H]
\caption{Best-fit 
 parameters and results for the fits described in Section~\ref{sec:likelihood1} using the Maximum Likelihood Estimator~\cite{Nielsen:2015nwa}. Including the bulk flow improves the quality of the fit and decreases the significance of accelerated expansion, as~seen from the decrease of $-2 \log {\cal L}_\textrm{max}$. ``No acceleration'' corresponds to $\Omega_\Lambda = \Omega_\mathrm{m}/2$.}
\label{tab:results3}\tablesize{\scriptsize}
\begin{adjustwidth}{-\extralength}{0cm}
\begin{tabularx}{\fulllength}
{c  c  c  c c  c  c  c  c  c  c   c  c  c }
\toprule
&\ \textbf{Fit} & \textbf{\boldmath{$-$}2 log\boldmath{${\cal L}_\textrm{max}$}} & \boldmath{$\Omega_\textrm{m}$} & \boldmath{$\Omega_\Lambda$} & \boldmath{$\alpha$} & \boldmath{$x_{1,0}$} & \boldmath{$\sigma_{x_{1,0}}$} & \boldmath{$\beta$} & \boldmath{$c_0$} & \boldmath{$\sigma_{c_0}$} & \boldmath{$M_0$} & \boldmath{$\sigma_{M_0}$} & \makecell[cc]{\textbf{\boldmath{$V$} (km s\boldmath{$^{-1}$}}) \\ \textbf{@\boldmath{$200h^{-1}$}Mpc}} \\ 
\midrule
\multirow{2}{*}{1.} &As in Ref.\cite{Nielsen:2015pga} & $-$214.97 & 0.341 & 0.569 & 0.134 & 0.0385 & 0.931 & 3.059 & $-$0.016 & 0.071 & $-$19.052 & 0.108 & -  \\  
 &No acceleration &$-$203.93 & 0.068 & 0.034 & 0.132 & 0.0327 & 0.932 & 3.045 & $-$0.013 & 0.071 & $-$19.006 & 0.110 & - \\
 \midrule
\multirow{2}{*}{2.} & \cite{Nielsen:2015pga} + JLA $z$ & $-$221.93 & 0.340 & 0.565 & 0.133 & 0.0385 & 0.932 & 3.056 & $-$0.016 & 0.071 & $-$19.051 & 0.107 & -  \\  
& No acceleration & $-$210.99 & 0.070 & 0.035 & 0.131 & 0.0328 & 0.932 & 3.042 &$-$0.013 & 0.071 & $-$19.006 & 0.109 & - \\
\midrule
\multirow{2}{*}{3.} & No pec. vel. corr. to $z$ & $-$215.40 & 0.285 & 0.483 & 0.134 & 0.0398 & 0.932 & 3.038 & $-$0.016 & 0.071 & $-$19.051 & 0.108 & - \\  
& No acceleration & $-$207.67 & 0.051 & 0.025 & 0.132 & 0.0348 & 0.932 & 3.023 & $-$0.014 & 0.071 & $-$19.012 & 0.110 & - \\
\midrule
\multirow{2}{*}{4.} & No pec. vel. corr. to $z$ or $m_B$ &$-$216.89 & 0.235 & 0.396 & 0.135 & 0.0397 & 0.932 & 3.029 & $-$0.016 & 0.071 & $-$19.040 & 0.109 & - \\  
& No acceleration & $-$211.84 & 0.0413 & 0.021 & 0.133 & 0.0357 & 0.932 & 3.016 & $-$0.014 & 0.071 & $-$19.008 & 0.110 & - \\
\midrule
\multirow{2}{*}{5.} & Exponential bulk flow &$-$217.51 & 0.289 & 0.452 & 0.134 & 0.0390 & 0.932 & 3.036 & $-$0.016 & 0.071 & $-$19.037 & 0.107 & 253 \\  
& No acceleration & $-$211.3 & 0.077 & 0.039 & 0.132 & 0.0347 & 0.932 & 3.024 & $-$0.014 & 0.071 & $-$19.002 & 0.108 & 292 \\
\midrule
\multirow{2}{*}{6.} & Linear bulk flow & $-$217.47 & 0.290 & 0.455 & 0.134 & 0.0390 & 0.932 & 3.036 & $-$0.016 & 0.071 & $-$19.038 & 0.107 & 265  \\  
& No acceleration & $-$211.99 & 0.082 & 0.041 & 0.132 & 0.0347 & 0.932 & 3.025 & $-$0.014 & 0.071 & $-$19.002 & 0.108 & 282  \\
\midrule
\multirow{2}{*}{7.} & JLA + Exp. bulk flow & $-$224.87 & 0.340 & 0.570 & 0.133 & 0.0387 & 0.932 & 3.051 & $-$0.016 & 0.072 & $-$19.052 & 0.107 & 271 \\  
& No acceleration & $-$216.3 & 0.077 & 0.039 & 0.132 & 0.0347 & 0.932 & 3.024 &$-$0.014 & 0.071 & $-$19.002 & 0.108 & 295 \\
\midrule
\multirow{2}{*}{8.} & JLA + Lin. bulk flow & $-$225.08 & 0.341 & 0.577 & 0.133 & 0.0387 & 0.932 & 3.050 & $-$0.016 & 0.071 & $-$19.054 & 0.107 & 238 \\  
& No acceleration & $-$214.14 & 0.072 & 0.036 & 0.131 & 0.0328 & 0.932 & 3.041 & $-$0.013 & 0.071 & $-$19.005 & 0.109 & 251 \\
\midrule
\multirow{3}{*}{9.} &  + Exp. Bulk Flow & $-$225.61 & 0.279 & 0.427 & 0.133 & 0.0386 & 0.932 & 3.001 &$-$0.016 & 0.071 & $-$19.034 & 0.109 & 309 \\  
& No acceleration & $-$220.72 & 0.086 & 0.043 & 0.132 & 0.0346 & 0.932 & 2.990 & $-$0.015 & 0.071 & $-$19.001 & 0.110 & 398 \\
& Flat & $-$223.96 & 0.393 & 0.607 & 0.133 & 0.0357 & 0.933 & 2.998 & $-$0.016 & 0.071 & $-$19.045 & 0.110 & 338 \\
\midrule
\multirow{3}{*}{10.} & + Lin. bulk flow & $-$225.73 & 0.277 & 0.431 & 0.133 & 0.0386 & 0.932 & 3.002 & $-$0.016 & 0.071 & $-$19.037 & 0.109 & 211 \\  
& No acceleration & $-$220.16 & 0.085 & 0.042 & 0.132 & 0.0346 & 0.932 & 2.991 & $-$0.015 & 0.071 & $-$19.001 & 0.110 & 249 \\
& Flat & $-$224.18 & 0.390 & 0.610 & 0.134 & 0.0399 & 0.932 & 3.006 & $-$0.016 & 0.071 &$-$19.047 & 0.109 & 215 \\
\bottomrule 
\end{tabularx}
\end{adjustwidth}
\end{table}

The results in Table~\ref{tab:results3} may be summarised as follows:

\begin{itemize}

\item Of all the fits, the~only ones favouring $\Omega_\Lambda > 0.5$ are just those that include the incorrect and incomplete peculiar velocity `corrections' of JLA \citep{Betoule:2014frx}.
  
\item Fit 4, \textls[-5]{which has no peculiar velocity corrections at all, as~in the cosmic acceleration discovery  papers~\cite{Perlmutter:1998np} and~\cite{Riess:1998cb}, prefers $\Omega_\Lambda = 0.396$ with $<$$2\sigma$ evidence for~acceleration.}

\item While previous work has suggested that bulk flows should not bias $\Omega_\Lambda$, it in fact drops by $\sim$$30\%$ if we undo the peculiar velocity `corrections' of JLA and instead use the kinematic data from . This illustrates the huge impact of considering a realistic LU-like observer such as ourselves, rather than the randomly located observer assumed in all previous analyses~\cite{Hui:2005nm,Neill:2007fh,Davis:2010jq,Huterer:2015gpa}. In~particular this contradicts what is stated in Table~11 of Ref.~\cite{Betoule:2014frx}.

\end{itemize}

The discovery papers~\cite{Perlmutter:1998np,Riess:1998cb} assumed the uncertainty due to peculiar velocities to be $c\sigma_z = 300$~km\,s$^{-1}$ and $200$~km\,s$^{-1}$ respectively, but~neither made SN-by-SN corrections. The~JLA \citep{Betoule:2014frx} and Pantheon \citep{Scolnic:2017caz} analyses  employ incorrect peculiar velocity `corrections', and adopt arbitrary redshift uncertainties of $c\sigma_z = 150$~km\,s$^{-1}$ and 250 km~s$^{-1}$, respectively.

\section{Extracting S\boldmath{$_8$}}
\label{sec:S8}
The peculiar velocity field 
 ${\bf v}({\bf r})$ is defined as
\begin{equation}
    {\bf v}({\bf r}) = \frac{H_0 f(\Omega_\textrm{m})}{4 \pi} \int d^3 {\bf r'}\delta({\bf r'})\frac{({\bf r'}- {\bf r})}{|{\bf r'}- {\bf r}|^3},
\end{equation}
where $f(\Omega_\textrm{m})$ is the logarithmic growth rate of fluctuations ($\simeq$$\Omega_\textrm{m}^{0.55}$ in the $\Lambda$CDM model) and $\delta({\bf r})$ is the matter density contrast field:
\begin{equation}
\delta({\bf r}) = \frac{\rho({\bf r}) - \bar{\rho}}{\bar{\rho}}.
\end{equation}
The density in the above equation is that of  gravitating matter, which is dominated by unobservable dark matter in the standard paradigm. It is usually assumed that observed luminous objects trace out the underlying matter density contrast with only linear bias
\begin{equation}
\delta_\textrm{g} = b_\textrm{g} \delta
\end{equation}
where $b_\textrm{g}$ and $\delta_\textrm{g}$ are the bias and the density fluctuation field of the tracers, respectively. The~predicted peculiar velocity field
\begin{equation}
    {\bf v}_\textrm{pred}({\bf r}) = \frac{H_0}{4 \pi} \int d^3 {\bf r'}\delta_\textrm{t}({\bf r'})\frac{({\bf r'}- {\bf r})}{|{\bf r'}- {\bf r}|^3}
\end{equation}
can now be compared to the  `observed' (in the case of the present study, simulated) to estimate the term $\beta_\textrm{t}$
\begin{equation}
\beta_\textrm{t} = \frac{{\bf v}_\textrm{t}}{{\bf v}_\textrm{pred}}
\end{equation}
for the tracer $\textrm{t}$ \citep{Hollinger:2021hwx, Boruah:2019icj}. Thus, $\beta_\textrm{t}$ can be obtained by fitting a large number of measured tracer velocities against the predictions from the density field. For~straightforward comparison with results from weak lensing, we convert this, as~in Ref.~\citep{Boruah:2019icj}, to~$S_8 \equiv \sigma_8 (\Omega_\textrm{m}/0.3)^{0.5}$, using the input value of $\Omega_\textrm{m} = 0.2952$ used for the \texttt{Dark Sky} simulations.

In practice, the~observational tracers used to map the velocity field need not be the same as are used to map the density field; in fact they usually are not~\citep{Boruah:2019icj}. However, since our aim is to study the effect of the local environment on velocity--density correlations, we use the same tracers for both, viz. the halos (we use halos rather than particles since the $z=0$ halo catalogue of \texttt{Dark Sky} is computationally more tractable than the $z=0$ particle snapshot). In order to keep our study as similar as possible to Ref.\citep{Hollinger:2021hwx} we use a Gaussian kernel with a smoothing length of 5 $h^{-1}$ Mpc to smooth out the density fluctuation~field.  

\begin{figure}[H]
    \includegraphics[width=0.7\columnwidth]{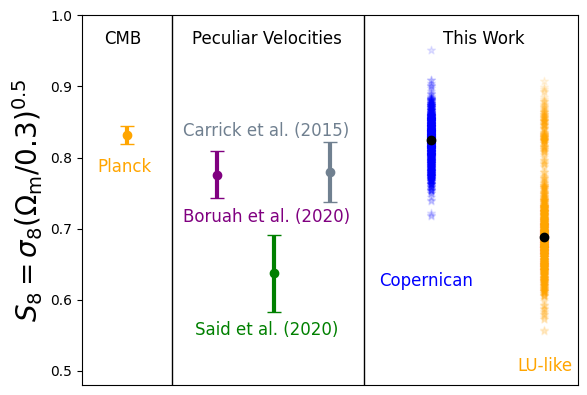}
    \caption{Distribution 
 of the $S_8$ parameter extracted by  comparing the `observed' (in this case, simulated) peculiar velocity with the prediction from the density contrast field, for~1000 observers selected at random (Copernican) as well as selected to be in Local Universe-like (LU-like) environments. The~median values are shown in black. For~comparison, the  measurements \citep{Carrick:2015xza,Boruah:2019icj,Said:2020epb} made using the same method are shown, as~well as the value derived from the~CMB measurements by Planck.}
    \label{fig:S8Summary}
\end{figure}

It is evident from Figure~\ref{fig:S8Summary} that there is a downward bias in measurements of $S_8$ by Local Universe-like observers such as ourselves and the cosmic variance is also higher. This may well account for the tension between the value derived from Planck data, and~that obtained by comparing the reconstructed velocity field from the 2M++ galaxy redshift compilation to supernovae, Fundamental Plane and  Tully–Fisher distances \citep{Carrick:2015xza,Boruah:2019icj,Said:2020epb}.

\section{Discussion}
\label{sec:discussion}

Other authors~\cite{Huterer:2015gpa,Huterer:2020szt}  have come to different conclusions regarding the impact of the bulk flow on cosmological parameter determination, so we comment on why this is so. 
In general relativity, space--time evolves according to the Einstein field equations, but for simplicity and tractability, structure formation is studied by linearising these around the maximally symmetric FLRW solution. By further making restricted `gauge' choices \citep{Bertschinger:1993xt}, 
cosmological reality is simulated using Newtonian N-body simulations wherein there is a background space that expands homogeneously and isotropically, governed by just one scale factor, while density perturbations evolve around this background according to Newtonian gravity. However, solutions to the linearised field equations can only be linearisations of the solutions to the fully non-linear equations \citep{Mukhanov:1990me,DEath:1976dwo}; hence, this approach hides  known physical phenomena. Indeed, a relativistic universe exhibits locally inhomogeneous expansion beyond that evident in linear perturbation theory around a maximally symmetric background \citep{Giblin:2015vwq}. It was therefore proposed to consider the expansion of the universe at late times as an average effect, arising out of the coarse-graining of physics at smaller scales \citep{Buchert:2015iva}. Peculiar velocities are thus at the interface between the real universe and its idealisation modelled by first-order perturbation of the fluid equations or in a Newtonian N-body simulation. Note that in the real universe, peculiar velocities are differences in the expansion rate of the universe at different space--time points, while in an N-body simulation they arise by construction from Newtonian gravity acting on top of a hypothetical uniformly expanding space. Refs.~\cite{Hui:2005nm, Huterer:2015gpa} use the~restricted longitudinal or conformal-Newtonian `gauge' \citep{Bertschinger:1993xt} to derive the covariance (Equation~\ref{eq:Sij}) for a typical observer. But,~as Figure~\ref{fig:DSvsTH} shows, this would be relevant for cosmology only if each SNe~Ia were being observed from a different, randomly sampled, host galaxy. In practice we observe the real universe from only one unique vantage point. Nevertheless, Ref.~\cite{Huterer:2015gpa} adds this covariance as a ``guaranteed theoretical signal'' to the uncertainty budget of the JLA data, thus weakening the preference for a bulk flow to $\lesssim$$2\sigma$ (as seen in their Figure~\ref{fig:lcdmlinearexponential}).

Ref.~\cite{Huterer:2020szt} claims further that any bias in the inference of dark energy parameters due to the effect of peculiar velocities can be determined {a priori} via simulations. This misses the point, however, that $\Lambda$CDM is a model and N-body simulations contain only as much physics as have been coded into them, i.e.,~neither capture the real universe. In fact Ref.~\cite{Huterer:2020szt} acknowledges that ignoring the velocity covariance altogether would lead to \emph{larger} effects due to  peculiar velocities---just as we have established here.

\section{Conclusions}
\label{sec:conclusion}
To summarise, we are \emph{not} typical (Copernican) observers--- we are embedded in a fast and deep bulk flow and this has significant impact on the covariances used in supernova cosmology
.\endnote{There are other corrections too such as for gravitational lensing, which become more important than the effect of peculiar velocities at redshift $z > 0.15$---see Figure B.1 of Ref.~\cite{Colin:2018ghy}.}
We have studied the effect on SNe~Ia  redshifts in the JLA catalogue of the peculiar velocities of their host galaxies. Using direct measurements of these from Cosmicflows-3, we find that the effect of peculiar velocities for low redshift SNe~Ia has been underestimated by 48\%. We show that the usual procedure of adding a constant velocity dispersion of a few hundred km~s$^{-1}$ to account for peculiar velocities at high redshift does not take into account the \emph{correlated} flow of the galaxies. By~analysing the \texttt{DarkSky} simulation \citep{Skillman:2014qca}, we demonstrate that `Local universe-like' observers like ourselves see a 2--8 times stronger correlation between the SNe~Ia than a randomly located observer does.
The JLA analysis \citep{Betoule:2014frx} corrected the data assuming the CMB dipole to be entirely kinematic in origin and that  convergence to the CMB rest frame occurs abruptly at redshift $z \sim 0.06$. Since neither assumption is fully supported by observations, we have adopted a general model of the bulk flow which introduces two additional parameters in the analysis. We do not adopt the $\Lambda$CDM model {a priori}, nor do we make assumptions about the origin of the CMB dipole. This provides an independent estimate of the bulk flow and we find that it persists out to distances beyond 200~$h^{-1}$Mpc, with~a speed of $\sim$$250$~km\,s$^{-1}$. Our maximum likelihood analysis then shows that the accelerated expansion of the universe cannot be inferred as a statistically significant result from the SNe~Ia data~alone.

\vspace{6pt} 



\authorcontributions{All authors 
 have contributed to all stages of this work, and read and agreed to the published version of the~manuscript.}

\funding{This research received no external~funding.}

\dataavailability{} 

\conflictsofinterest{The authors declare no conflicts of~interest.} 
\appendixtitles{yes} 
\appendixstart
\appendix
\section[\appendixname~\thesection]{Supernova Cosmology}
\label{SNcosmo}

The largest public catalogues of SNe~Ia lightcurves, the~JLA \citep{Betoule:2014frx} and its successor Pantheon \citep{Scolnic:2017caz}, employ the `Spectral Adaptive Lightcurve Template 2' (SALT2) to fit each SNe~Ia light curve with 3 parameters: the apparent magnitude $m^*_B$ (at maximum in the rest frame `$B$-band') and the `shape' and `colour' corrections, $x_1$ and $c$ \citep{Guy:2007dv} (a `host galaxy mass correction' may also be included, however the MLE is insensitive to this parameter~\cite{Nielsen:2015nwa}). 
The distance modulus is then
\begin{equation}
 \label{eq:DMOD}   
\mu_\mathrm{SN} = m^*_B - M + \alpha x_1 - \beta c, 
\end{equation}
where $\alpha$ and $\beta$ are assumed to be constants, as~is $M$ the absolute SNe~Ia magnitude, as~befits a `standard candle'. 
In the
standard $\Lambda$CDM cosmological model, this is related to the luminosity distance $d_\textrm{L}$ as
\begin{eqnarray}
& \mu &\equiv 25 + 5 \log_{10}(d_\textrm{L}/\textrm{Mpc}), 
 \quad \textrm{where:} \nonumber \\
& d_\textrm{L} &= (1 + z) \frac{d_\textrm{H}}{\sqrt{\Omega_k}} 
 \textrm{sin}\left(\sqrt{\Omega_k} \int_0^z \frac{H_0 \textrm{d}z'}{H(z')}\right), \textrm{for } \Omega_k > 0
 \nonumber \\
&  &= (1 + z) d_\textrm{H} \int_0^z \frac{H_0 \textrm{d}z'}{H(z')}, \textrm{for } \Omega_k = 0
 \nonumber \\
&  &= (1 + z) \frac{d_\textrm{H}}{\sqrt{\Omega_k}} 
 \textrm{sinh}\left(\sqrt{\Omega_k} \int_0^z \frac{H_0 \textrm{d}z'}{H(z')}\right), \textrm{for } \Omega_k < 0
 \nonumber \\
& d_\textrm{H} &= c/H_0 \simeq 3000h^{-1}~\textrm{Mpc}, H_0 \equiv 
 100h~\textrm{km}\,\textrm{s}^{-1}\textrm{Mpc}^{-1}, \nonumber \\
& H &= H_0 \sqrt{\Omega_\textrm{m} (1 + z)^3 + \Omega_k (1 + z)^2 
 + \Omega_\Lambda}.
\label{eq:DLEQ}
\end{eqnarray}
Here, $H$, the Hubble parameter ($H_0$ being its present value), $d_\textrm{H}$ is the `Hubble distance' and $\Omega_\textrm{m}, \Omega_\Lambda, \Omega_k$ are the matter, cosmological constant and curvature densities in units of the critical density. In~the standard $\Lambda$CDM model, these are related by the `cosmic sum rule': $1=\Omega_\textrm{m} + \Omega_\Lambda + \Omega_k$, which is simply rewriting the Friedmann~equation. 

Thus, knowing the redshift and distance of the `standardised' SNe~Ia, one can determine the cosmological parameters. However these are measured from Earth and usually quoted in the heliocentric frame (allowing for the Earth's motion around the Sun), so they need to first be translated to the reference frame in which the universe is (statistically) isotropic and homogeneous and the above equations hold. Assuming that the CMB dipole is purely kinematic in origin, this is taken to be the `CMB frame' which can be reached by a local special relativistic boost, so the measured values are corrected as  in Equation~(\ref{eq:zt}) for the redshift and Equation~(\ref{eq:dlt}) for the luminosity~distance.

\section{The Joint Lightcurve Analysis~Catalogue}
\label{JLA}

\begin{figure}[H]
    \includegraphics[width=0.8\columnwidth]{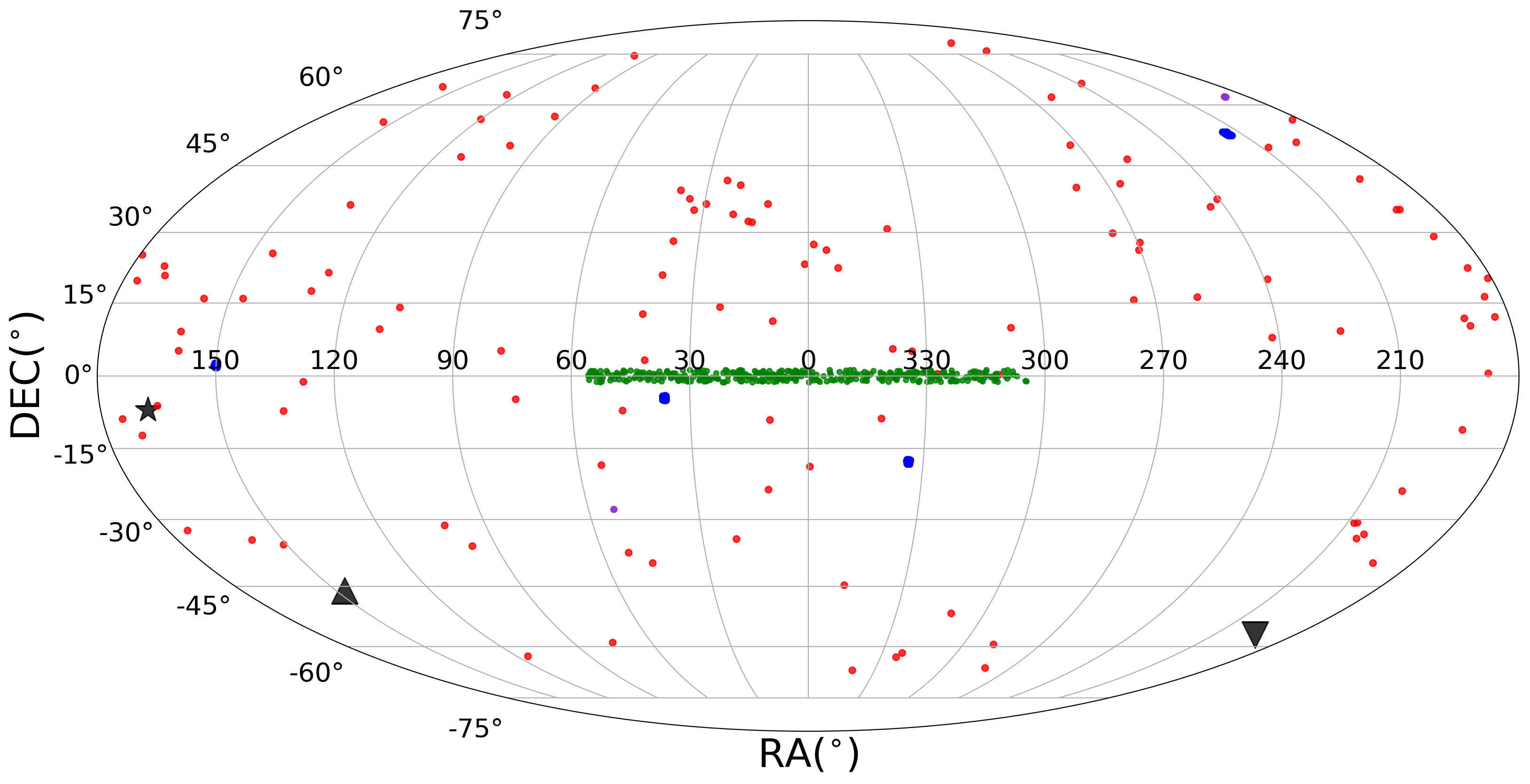}
    \caption{Sky distribution (Mollweide projection, equatorial coordinates) of the 4 subsamples of the JLA catalogue: low $z$ (red dots), SDSS (green dots), HST (black dots), clusters of many SNe~Ia from SNLS (blue dots) \. The~directions of the CMB dipole (star), the~SMAC bulk flow (triangle) \protect\citep{Neill:2007fh}, and~the 2M++ bulk flow (inverted triangle)  \protect\citep{Carrick:2015xza} are shown in~grey.}
    \label{fig:JLASkyScatter}
\end{figure}

\begin{figure}[H]
    \includegraphics[width=0.6\columnwidth]{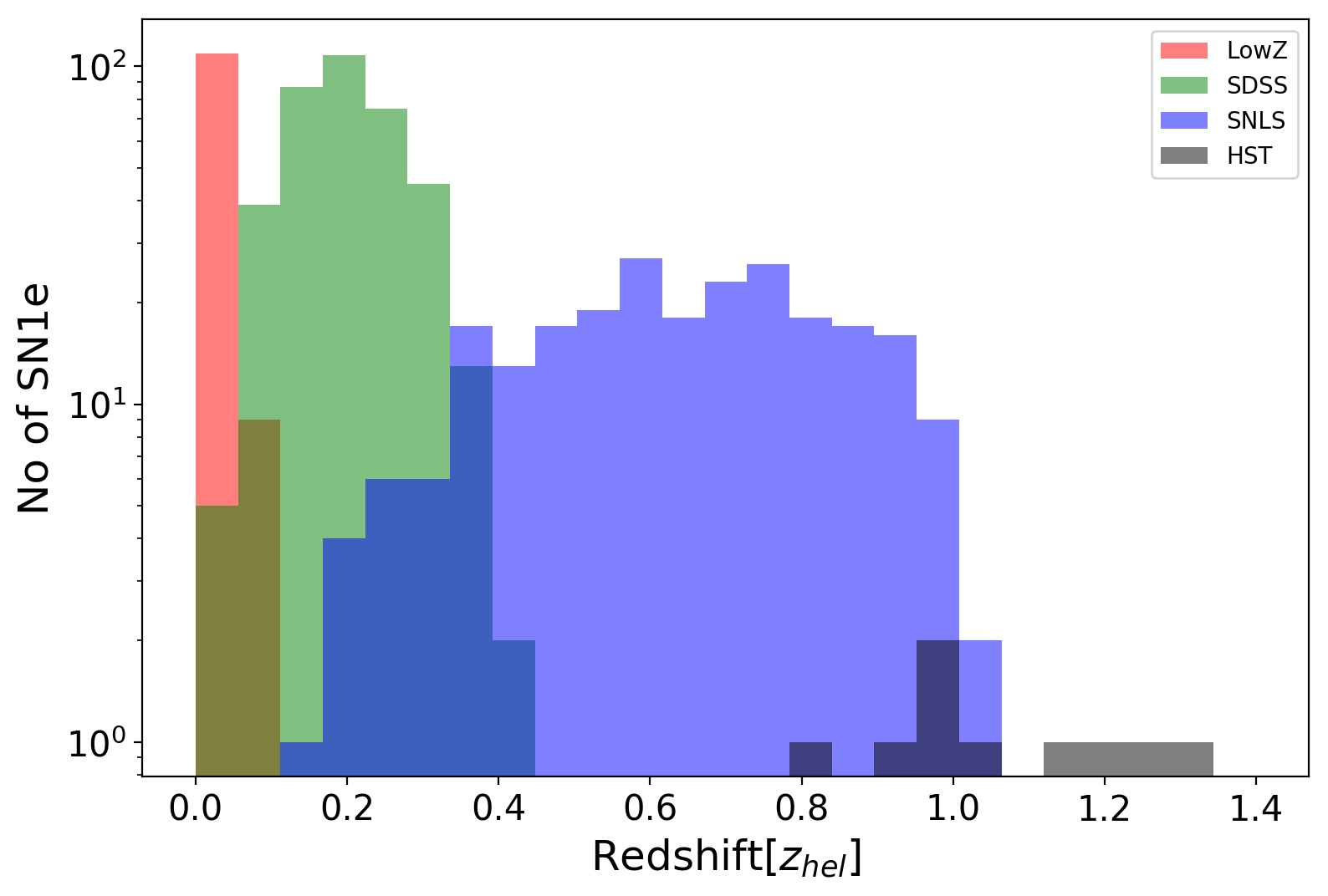}
    \caption{The redshift distribution of the 4 samples that make up the SDSS-II/SNLS3 Joint Lightcurve Analysis~catalogue.}
    \label{fig:JLAZDist}
\end{figure}

The JLA catalogue \citep{Betoule:2014frx} consists of 740 spectroscopically confirmed SNe~Ia, including several low redshift ($z<0.1$) samples, three seasons of SDSS-II ($0.05 < z < 0.4$) and three years of SNLS ($0.2<z<1)$ data, all calibrated consistently in the `Spectral Adaptive Lightcurve Template 2' (SALT2) scheme. Figures~\ref{fig:JLASkyScatter} and \ref{fig:JLAZDist} show, respectively, the sky coverage and redshift distribution of this publicly available~catalogue.

\unskip

We use the publicly available SALT2 light curve fits carried out by the JLA collaboration~\citep{Betoule:2014frx}, but rather than use their `constrained' $\chi^2$ statistic which is unprincipled, we employ the Maximum Likelihood Estimator (MLE) due to \citet{Nielsen:2015pga}.  
Note that the distribution of `pulls' using this MLE is gaussian, while the best fit and its uncertainties explicitly satisfy Wilk's theorem \citep{Nielsen:2015pga}. 
Our approach is frequentist but equivalent to the `Bayesian Hierarchical Model'  \citep{March:2011xa, Shariff:2015yoa, Rubin:2016iqe}. It has been used in independent analyses of SNe~Ia~\cite{Dam:2017xqs, Desgrange:2019npu}.

Ref.~\cite{Rubin:2016iqe} advocated that the shape and colour parameters be allowed to depend on both the SNe~Ia sample and the redshift; however, this introduces 12 additional parameters (to the 10 used above) and thus violates the Bayesian information criterion~\citep{Colin:2018ghy, Colin:2019ulu}. 
Ref.~\cite{Karpenka:2015vva} noted that if $x_1$ and $c$ evolve with redshift, the~likelihood-based methods return biased values of the parameters (while the `constrained $\chi^2$' method continues to be robust); however, this conclusion is arrived at using Monte Carlo simulations which assume the $\Lambda$CDM model and is therefore a circular argument. It has been emphasised by~\cite{Dam:2017xqs} that systematic uncertainties and selection biases in the data need to be corrected for in a model-independent manner, before fitting to a particular cosmological model. For~further discussion of these and other related issues, see Ref.~\cite{Mohayaee:2021jzi}.


\begin{adjustwidth}{-\extralength}{0cm}
\printendnotes[custom]
\reftitle{References}



\begin{thebibliography}{999}

%



\bibitem{Peebles:2022akh}
Peebles, P.J.E.\ 
{Anomalies in physical cosmology}.
{\textit{Ann. Phys.} \textbf{2022}, \textit{447}, 169159}.
\url{https://doi.org/10.1016/j.aop.2022.169159}.


\bibitem{Freedman:2021ahq}
Freedman, W.L.\ 
{Measurements of the Hubble constant: Tensions in perspective}
{\textit{Astrophys. J.} \textbf{2021}, \textit{919}, 16}.
\url{https://doi.org/10.3847/1538-4357/ac0e95}.


\bibitem{Abdalla:2022yfr}
 Abdalla, E.; Franco, Abell\'an, G.; Aboubrahim, A.; Agnello, A.; Akarsu, O.; Akrami, Y.; Alestas, G.; Aloni, D.; Amendola, L.; Anchordoqui, L.A.; 
 {et al.}\ 
 {Cosmology intertwined: A review of the particle physics, astrophysics, and cosmology associated with the cosmological tensions and anomalies}.
 {\em J. High Energy Astropys.} \textbf{2022}, \textit{34}, 49. \url{https://doi.org/10.1016/j.jheap.2022.04.002}.


\bibitem[Secrest et al.(2021)]{Secrest:2020has}
Secrest, N.J.; von Hausegger, S.; Rameez, M.; Mohayaee, R.; Sarkar, S.; Colin, J.\ 
{A Test of the Cosmological Principle with Quasars}.
{\textit{Astrophys. J. Lett.} \textbf{2021}, \textit{908}, L51}.
\url{https://doi.org/10.3847/2041-8213/abdd40}.


\bibitem[Secrest et al.(2022)]{Secrest:2022uvx}
Secrest, N.J.; von Hausegger, S.; Rameez, M.; Mohayaee, R.; Sarkar, S.\ 
{A Challenge to the Standard Cosmological Model}.
{\textit{Astrophys. J. Lett.} \textbf{2022}, \textit{937}, L31}.
\url{https://doi.org/10.3847/2041-8213/ac88c0}.


\bibitem[Colin et al.(2019a)]{Colin:2018ghy}
 Colin, J.; Mohayaee, R.; Rameez, M.; Sarkar, S.\ 
 {Evidence for anisotropy of cosmic acceleration}.
 {\textit{Astron. Astrophys.} \textbf{2019}, \textit{631}, L13}.
 \url{https://doi.org/10.1051/0004-6361/201936373}.
 

\bibitem[Tsagas et al.(2021)]{Tsagas:2021dsl}
Tsagas, C.G.; Kadiltzoglou, M.I.; Asvesta, K.\ 
{The deceleration parameter in \textquotedblleft{}tilted\textquotedblright{} Friedmann universes: Newtonian vs relativistic treatment}.
{\textit{Astrophys. Space Sci.} \textbf{2021}, \textit{366}, 90}.
\url{https://doi.org/10.1007/s10509-021-03995-7}.


\bibitem[Watkins et al.(2023)]{Watkins:2023rll}
Watkins, R.; Allen, T.; Bradford, C.J.; Walker, A.; Feldman, H.A.; Cionitti, R.; Al-Shorman, Y.; Kourkchi, E.; Tully, R.B.\ 
{Analysing the large-scale bulk flow using Cosmicflows4: Increasing tension with the standard cosmological model}.
{\textit{Mon. Not. Roy. Astron. Soc.} \textbf{2023}, \textit{524}, 1885}.
\url{https://doi.org/10.1093/mnras/stad1984}.


\bibitem[McClure \& Dyer (2007)]{McClure:2007vv}
McClure, M.L.; Dyer, C.C.\ 
{Anisotropy in the Hubble constant as observed in the HST Extragalactic Distance Scale Key Project results}.
{\textit{New Astron.} \textbf{2007}, \textit{12}, 533}.
\url{https://doi.org/10.1016/j.newast.2007.03.005}.


\bibitem[Leibundgut(2000)]{Leibundgut:2000xw}
 Leibundgut, B.\ 
~{Type Ia supernovae}.
{\textit{Astron. Astrophys. Rev.} \textbf{2000}, \textit{10}, 179}.
\url{https://doi.org/10.1007/s001590000009}.


\bibitem{Brout:2022vxf}
Brout, D.; Scolnic, D.; Popovic, B.; Riess, A.G.; Zuntz, J.; Kessler, R.; Carr, A.; Davis, T.M.; Hinton, S.; Jones, D.; {et al.}\ 
{The Pantheon+ Analysis: Cosmological Constraints}.
{\textit{Astrophys. J.} \textbf{2022}, \textit{938}, 110}.
\url{https://doi.org/10.3847/1538-4357/ac8e04}.


\bibitem[Perlmutter et al.(1999)]{Perlmutter:1998np}
Perlmutter, S.; Aldering, G.; Goldhaber, G.; Knop, R.A.; Nugent, P.; Castro, P.G.; Deustua, S.; Fabbro, S.; Goobar, A.; Groom, D.E.; {et al.}\ 
 {Measurements of Omega and Lambda from 42 high redshift supernovae}.
 {\textit{Astrophys. J.} \textbf{1999}, \textit{517}, 565}.
 \url{https://doi.org/10.1086/307221}.


\bibitem[Riess et al.(1998)]{Riess:1998cb}
 Riess, A.G.; Filippenko, A.V.; Challis, P.; Clocchiatti, A.; Diercks, A.; Garnavich, P.M.; Gillil, ; R.L.; Hogan, C.J.; Jha, S.; Kirshner, R.P.; {et al.}\ 
 {Observational evidence from supernovae for an accelerating universe and a cosmological constant}.
 {\textit{Astron. J.} \textbf{1998}, \textit{116}, 1009}.
 \url{https://doi.org/10.1086/300499}.


\bibitem[Hudson et al.(2004)]{Hudson:2004et}
 Hudson, M.J.; Smith, R.J.; Lucey, J.R.; Branchini, E.\ 
 {Streaming motions of galaxy clusters within 12000 km/s. 5. The peculiar velocity field}.
 {\textit{Mon. Not. Roy. Astron. Soc.} \textbf{2004}, \textit{352}, 61}.
 \url{https://doi.org/10.1111/j.1365-2966.2004.07893.x}.
 

\bibitem[Conley et al.(2011)]{Conley:2011ku} 
 Conley A.; Guy J.; Sullivan M.; Regnault, N.; Astier, P.; Ball, C.; Basa, S.; Carlberg, R.G.; Fouchez, D.; Hardin, D.; et al.\ 
 {Supernova Constraints and Systematic Uncertainties from the First 3 Years of the Supernova Legacy Survey}.
 {\textit{Astrophys. J. Suppl.} \textbf{2011}, \textit{192}, 1}.
 \url{https://doi.org/10.1088/0067-0049/192/1/1}.
 

\bibitem[Neill et al.(2007)]{Neill:2007fh}
 Neill, J.D.; Hudson, M.J.; Conley, A.\ 
 {The peculiar velocities of local Type Ia supernovae and their impact on cosmology}.
 {\textit{Astrophys. J. Lett.} \textbf{2007}, \textit{661}, L123}.
 \url{https://doi.org/10.1086/518808}.


\bibitem[Saunders et al.(2000)]{Saunders:2000af}
 Saunders, W.; Sutherl, W.J.; Maddox, S.J.; Keeble, O.; Oliver, S.J.; Rowan-Robinson, M.; McMahon, R.G.; Efstathiou, G.P.; Tadros, H.; White, S.D.; {et al.}\ 
 {The PSCz catalogue}.
 {\textit{Mon. Not. Roy. Astron. Soc.} \textbf{2000}, \textit{317}, 55}.
 \url{https://doi.org/10.1046/j.1365-8711.2000.03528.x}.


\bibitem[Carrick et al.(2015)]{Carrick:2015xza}
 Carrick, J.; Turnbull, S.J.; Lavaux, G.; Hudson, M.J.\ 
 {Cosmological parameters from the comparison of peculiar velocities with predictions from the 2M++ density field}.
 {\textit{Mon. Not. Roy. Astron. Soc.} \textbf{2015}, \textit{450}, 317}.
 \url{https://doi.org/10.1093/mnras/stv547}


\bibitem[Colin et al.(2011)]{Colin:2010ds}
 Colin, J.; Mohayaee, R.; Sarkar, S.; Shafieloo, A.\ 
 {Probing the anisotropic local universe and beyond with SNe~Ia data}. 
 {\textit{Mon. Not. Roy. Astron. Soc.} \textbf{2011}, \textit{414}, 264}.
 \url{https://doi.org/10.1111/j.1365-2966.2011.18402.x}.


\bibitem[Feindt et al.(2013)]{Feindt:2013pma}
 Feindt, U.; Kerschhaggl, M.; Kowalski, M.; Aldering, G.; Antilogus, P.; Aragon, C.; Bailey, S.; Baltay, C.; Bongard, S.; Buton, C.; {et al.}\ 
 {Measuring cosmic bulk flows with Type Ia supernovae from the Nearby Supernova Factory}.
 {\textit{Astron. Astrophys.} \textbf{2013}, \textit{560}, A90}.
 \url{https://doi.org/10.1051/0004-6361/201321880}.


\bibitem[Howlett et al.(2022)]{Howlett:2022len}
Howlett, C.; Said, K.; Lucey, J.R.; Colless, M.; Qin, F.; Lai, Y.; Tully, R.B.; Davis, T.M.\ 
{The sloan digital sky survey peculiar velocity catalogue}.
{\textit{Mon. Not. Roy. Astron. Soc.} \textbf{2022}, \textit{515}, 953}.
\url{https://doi.org/doi:10.1093/mnras/stac1681}.


\bibitem[Lavaux et al.(2010)]{Lavaux:2008th}
 Lavaux, G.; Tully, R.B.; Mohayaee, R.; Colombi, S.\  
 {Cosmic flow from 2MASS redshift survey: The origin of CMB dipole and implications for LCDM cosmology}.
 {\textit{Astrophys. J.} \textbf{2010}, \textit{709}, 48}.
 \url{https://doi.org/10.1088/0004-637X/709/1/483}.


\bibitem[Magoulas et al.(2016)]{2016IAUS..308..336M}
 Magoulas, C.; Springob, C.; Colless, M.; Mould, J.; Lucey, J.; Erdo{\v, g}du, P.; Jones, D.H.\ 
 {Measuring the cosmic bulk flow with 6dFGSv}.
 {\textit{Proc. IAU Symp.} \textbf{2016}, \textit{308}, 336}.
 \url{https://doi.org/10.1017/S1743921316010115}.


\bibitem[Watkins et al.(2009)]{Watkins:2008hf}
 Watkins, R.; Feldman, H.A.; Hudson, M.J.\ 
 {Large-scale bulk flows from the Cosmicflows-2 catalogue}.
 {\textit{Mon. Not. Roy. Astron. Soc.} \textbf{2009}, \textit{392}, 743}.
 \url{https://doi.org/10.1111/j.1365-2966.2008.14089.x}.


\bibitem[Betoule et al.(2014)]{Betoule:2014frx}
Betoule, M.; Kessler, R.; Guy, J.; Mosher, J.; Hardin, D.; Biswas, R.; Astier, P.; El-Hage, P.; Konig, M.; Kuhlmann, S.; {et al.}\ 
 {Improved cosmological constraints from a joint analysis of the SDSS-II and SNLS supernova samples}.
{\textit{Astron. Astrophys.} \textbf{2014}, \textit{568}, A22}.
 \url{https://doi.org/10.1051/0004-6361/201423413}.


\bibitem[Scolnic et al. (2017)]{Scolnic:2017caz}
 Scolnic, D.M.; Jones, D.O.; Rest, A.; Pan, Y.C.; Chornock, R.; Foley, R.J.; Huber, M.E.; Kessler, R.; Narayan, G.; Riess, A.G.; {et al.}\ 
 {The Complete Light-curve Sample of Spectroscopically Confirmed SNe~Ia from Pan-STARRS1 and Cosmological Constraints from the Combined Pantheon Sample}.
 {\textit{Astrophys. J.} \textbf{2018}, \textit{859}, 101}.
 \url{https://doi.org/doi.org/10.3847/1538-4357/aab9bb}.


\bibitem[Rameez \& Sarkar (2021)]{Rameez:2019wdt}
Rameez, M.; Sarkar, S.\ 
 {Is there really a Hubble tension?}
 {\textit{Class. Quant. Grav.} \textbf{2021}, \textit{38}, 154005}.
 \url{https://doi.org/10.1088/1361-6382/ac0f39}.


\bibitem[Davis et al.(2011)]{Davis:2010jq}
 Davis, T.M.; Hui, L.; Frieman, J.A.; Haugb{\o}lle, T.; Kessler, R.; Sinclair, B.; Sollerman, J.; Bassett, B.; Marriner, J.; M{\o}rtsell, E.; et al.\ 
 {The effect of peculiar velocities on supernova cosmology}.
 {Astrophys. J. \textbf{2011}, \textit{741}, 67}.
 \url{https://doi.org/10.1088/0004-637X/741/1/67}.


\bibitem[Hui \& Greene(2006)]{Hui:2005nm}
 Hui, L.; Greene, P.B.\ 
 {Correlated fluctuations in luminosity distance and the (surprising) importance of peculiar motion in supernova surveys}.
 {\textit{Phys. Rev. D} \textbf{2006}, \textit{73}, 123526}.
 \url{https://doi.org/10.1103/PhysRevD.73.123526}.


\bibitem[Huterer et~al.(2015)]{Huterer:2015gpa} 
 Huterer, D.; Shafer, D.L.; Schmidt, F.\ 
 {No evidence for bulk velocity from Type Ia supernovae}.
 {\textit{J. Cosmo. Astropart. Phys.} \textbf{2015}, \textit{12}, 033}.
 \url{https://doi.org/10.1088/1475-7516/2015/12/033}.
 

\bibitem[Hunt \& Sarkar(2010)]{Hunt:2008wp}
 Hunt, P.; Sarkar, S.\ 
 {Constraints on large scale inhomogeneities from WMAP-5 and SDSS: Confrontation with recent observations}.
 {\textit{Mon. Not. Roy. Astron. Soc.} \textbf{2010}, \textit{401}, 547}.
 \url{https://doi.org/10.1111/j.1365-2966.2009.15670.x}.
  

\bibitem[Hellwing et al.(2017)]{Hellwing:2016pdl}
 Hellwing, W.A.; Nusser, A.; Feix, M.; Bilicki, M.\ 
 {Not a Copernican observer: Biased peculiar velocity statistics in the local universe}.
 {\textit{Mon. Not. Roy. Astron. Soc.} \textbf{2017}, \textit{467}, 2787}.
 \url{https://doi.org/10.1093/mnras/stx213}.
 

\bibitem[Hellwing et al.(2018)]{Hellwing:2018tiq}
 Hellwing, W.A.; Bilicki, M.; Libeskind, N.I.\ 
 {Uneven flows: On cosmic bulk flows, local observers, and gravity}.
 {\textit{Phys. Rev. D} \textbf{2018}, \textit{97}, 103519}.
 \url{https://doi.org/10.1103/PhysRevD.97.103519}.
 

\bibitem[Rameez et al.(2018)]{Rameez:2017euv}
 Rameez, M.; Mohayaee, R.; Sarkar, S.; Colin, J.\ 
 {The dipole anisotropy of AllWISE galaxies}.
 {\textit{Mon. Not. Roy. Astron. Soc.} \textbf{2018}, \textit{477}, 1772}.
 \url{https://doi.org/10.1093/mnras/sty619}.


\bibitem[Skillman et al.(2014)]{Skillman:2014qca}
 Skillman, S.W.; Warren, M.S.; Turk, M.J.; Wechsler, R.H.; Holz, D.E.; Sutter, P.M.\ 
 {Dark Sky simulations: Early data release}.
 \emph{arXiv} \textbf{2014}, {arXiv:1407.2600}.


\bibitem[Tully et al.(2016)]{Tully:2016ppz} 
 Tully, R.B.; Courtois, H.M.; Sorce, J.G.\ 
 {Cosmicflows-3}.
{\textit{Astron J.} \textbf{2016}, \textit{152}, 50}.
\url{https://doi.org/10.3847/0004-6256/152/2/50}.


\bibitem[Nielsen et al.(2016)]{Nielsen:2015pga}
 Nielsen, J.T.; Guffanti, A.; Sarkar, S.\ 
 {Marginal evidence for cosmic acceleration from Type Ia supernovae}.
 {\textit{Sci. Rep.} \textbf{2016}, \textit{6}, 35596}. 
 \url{https://doi.org/10.1038/srep35596}.


\bibitem[Rubin \& Hayden(2016)]{Rubin:2016iqe}
  Rubin, D.; Hayden, B.\ 
  {Is the expansion of the universe accelerating? All signs point to yes}.
  {\textit{Astrophys. J. Lett.} \textbf{2016}, \textit{833}, L30}.
  \url{https://doi.org/10.3847/2041-8213/833/2/L30}.


\bibitem[Boruah et al.(2020)]{Boruah:2019icj}
Boruah, S.S.; Hudson, M.J.; Lavaux, G.\ 
Cosmic flows in the nearby universe: New peculiar velocities from SNe and cosmological constraints.
{\textit{Mon. Not. Roy. Astron. Soc.} \textbf{2020}, \textit{498}, 2703}.
\url{https://doi.org/10.1093/mnras/staa2485}


\bibitem[Cautun et al.(2014)]{Cautun:2014dda}
 Cautun, M.; Frenk, C.S.; van de Weygaert, R.; Hellwing, W.A.; Jones, B.J.T.\ 
 {Milky Way mass constraints from the Galactic satellite gap}.
 {\textit{Mon. Not. Roy. Astron. Soc.} \textbf{2014}, \textit{445}, 2049}.
 \url{https://doi.org/10.1093/mnras/stu1849}.


\bibitem[Hong et al.(2014)]{Hong:2014jla}
 Hong, T.; Springob, C.M.; Staveley-Smith, L.; Scrimgeour, M.I.; Masters, K.L.; Macri, L.M.; Koribalski, B.S.; Jones, D.H.; Jarrett, T.H.\ 
{2MTF \textendash{} IV. A bulk flow measurement of the local universe}.
{\textit{Mon. Not. Roy. Astron. Soc.} \textbf{2014}, \textit{445}, 402}.
\url{https://doi.org/10.1093/mnras/stu1774}.


\bibitem[Boggs et al.(1990)]{ODR}
 Boggs, P.T.; Rogers, J.E.\ 
 {Orthogonal distance regression}.
 {\textit{Contemp. Math.} \textbf{1990}, \textit{112}, 183}.
 \url{http://dx.doi.org/10.1090/conm/112}.


\bibitem[Colin et al.(2017)]{Colin:2017juj}
 Colin, J.; Mohayaee, R.; Rameez, M.; Sarkar, S.\ 
 {High redshift radio galaxies and divergence from the CMB dipole}.
 {\textit{Mon. Not. Roy. Astron. Soc.} \textbf{2017}, \textit{471}, 1045}.
 \url{https://doi.org/10.1093/mnras/stx1631}.


\bibitem[Hollinger et al.(2021)]{Hollinger:2021hwx}
Hollinger, A.M.; Hudson, M.J.\ 
Assessing the accuracy of cosmological parameters estimated from velocity \textendash{} density comparisons via simulations.
{\textit{Mon. Not. Roy. Astron. Soc.} \textbf{2021}, \textit{502}, 3723}.
\url{https://doi.org/10.1093/mnras/staa4039}.


\bibitem{Said:2020epb}
Said, K.; Colless, M.; Magoulas, C.; Lucey, J.R.; Hudson, M.J.\ 
{Joint analysis of 6dFGS and SDSS peculiar velocities for the growth rate of cosmic structure and tests of gravity}.
{\textit{Mon. Not. Roy. Astron. Soc.} \textbf{2020}, \textit{497}, 1275}.
\url{https://doi.org/10.1093/mnras/staa2032}.


\bibitem[Huterer (2020)]{Huterer:2020szt}
 Huterer, D.\ 
 {Specific effect of peculiar velocities on dark-energy constraints from Type Ia supernovae}.
 {\textit{Astrophys. J. Lett.} \textbf{2020}, \textit{904}, L28}.
 \url{https://doi.org/10.3847/2041-8213/abc958}.
 

\bibitem[Bertschinger (1993)]{Bertschinger:1993xt}
  Bertschinger, E.\ 
  {Cosmological dynamics: Course 1}. \emph{arXiv}  \textbf{1993},
  arXiv:astro-ph/9503125.


\bibitem[Mukhanov et al.(1990)]{Mukhanov:1990me}
 Mukhanov, V.F.; Feldman, H.A.; ; Br, enberger, R.H.\ 
 {Theory of cosmological perturbations. Part 1. Classical perturbations. Part 2. Quantum theory of  perturbations. Part 3. Extensions},
 {\textit{Phys. Rep.} \textbf{1992}, \textit{215}, 203}.
 \url{https://doi.org/10.1016/0370-1573(92)90044-Z}.


\bibitem[D'Eath (1976)]{DEath:1976dwo}
 D'Eath, P.D.\ 
 {On the existence of perturbed Robertson-Walker universes}.
{\textit{Ann. Phys.} \textbf{1976}, \textit{98}, 237}.  \url{https://doi.org/10.1016/0003-4916(76)90246-3}.


\bibitem[Giblin et al.(2015)]{Giblin:2015vwq}
 Giblin, J.T.; Mertens, J.B.; Starkman, G.D.\ 
 {Departures from the Friedmann-Lemaitre-Robertston-Walker cosmological model in an inhomogeneous universe: A numerical examination}.
 {\textit{Phys. Rev. Lett.} \textbf{2015}, \textit{116}, 251301}.
 \url{https://doi.org/10.1103/PhysRevLett.116.251301}.


\bibitem{Buchert:2015iva}
Buchert, T.; Carfora, M.; Ellis, G.F.R.; Kolb, E.W.; MacCallum, M.A.H.; Ostrowski, J.J.; R\"as\"anen, S.; Roukema, B.F.; Andersson, L.; Coley, A.A.; {et al.}\  
{Is there proof that backreaction of inhomogeneities is irrelevant in cosmology?}
{\textit{Class. Quant. Grav.} \textbf{2015}, \textit{32}, 215021}.
\url{https://doi.org/10.1088/0264-9381/32/21/215021}.
 

\bibitem[Guy et al.(2007)]{Guy:2007dv}
 Guy, J.; Astier, P.; Baumont, S.; Hardin, D.; Pain, R.; Regnault, N.; Basa, S.; Carlberg, R.G.; Conley, A.; Fabbro, S.; {et al.}\ 
 {SALT2: Using distant supernovae to improve the use of Type Ia supernovae as distance indicators}.
 {\textit{Astron. Astrophys.} \textbf{2007}, \textit{466}, 11}.
 \url{https://doi.org/10.1051/0004-6361:20066930}.


\bibitem[Nielsen(2015)]{Nielsen:2015nwa}
 Nielsen, J.T.\ 
 {Supernovae as cosmological probes}. 
 \emph{arXiv} \textbf{2015}, arXiv:1508.07850.


\bibitem[March et al.(2011)]{March:2011xa}
 March, M.; Trotta, R.; Berkes, P.; Starkman, G.; Vaudrevange, P.\ 
 Improved constraints on cosmological parameters from Type Ia supernova data. 
 {\textit{Mon. Not. Roy. Astron. Soc.} \textbf{2011}, \textit{418}, 2308--2329}.
 \url{https://doi.org/10.1111/j.1365-2966.2011.19584.x}.


\bibitem[Shariff et al.(2015)]{Shariff:2015yoa}
 Shariff, H.; Xiao, J.; Trotta, R.; van Dyk, D.A.\ 
 {BAHAMAS: New analysis of Type Ia supernovae reveals inconsistencies with standard cosmology}.
 {\textit{Astrophys. J.} \textbf{2016}, \textit{827}, 1}.
 \url{https://doi.org/10.3847/0004-637X/827/1/1}.


\bibitem[Dam, Heinesen \& Wiltshire(2017)]{Dam:2017xqs}
 Dam, L.H.; Heinesen, A.; Wiltshire, D.L.\ 
 {Apparent cosmic acceleration from Type Ia supernovae}.
 {\textit{Mon. Not. Roy. Astron. Soc.} \textbf{2017}, \textit{472}, 835}.
 \url{https://doi.org/10.1093/mnras/stx1858}.
 

\bibitem[Desgrange et al.(2019)]{Desgrange:2019npu}
 Desgrange, C.; Heinesen, A.; Buchert, T.\ 
 {Dynamical spatial curvature as a fit to Type Ia supernovae}
 {\textit{Intern. J. Mod. Phys. D} \textbf{2019}, \textit{28}, 1950143}.
 \url{https://doi.org/10.1142/S0218271819501438}.
  

\bibitem[Colin et al.(2019b)]{Colin:2019ulu}
 Colin, J.; Mohayaee, R.; Rameez, M.; Sarkar, S.\ 
 {A response to Rubin \& Heitlauf: ``Is the expansion of the universe accelerating? All signs still point to yes''}.
 \emph{arXiv} \textbf{2019}, arXiv:1912.04257.
 

\bibitem[Karpenka(2015)]{Karpenka:2015vva}
 Karpenka, N.V.\ 
 {The supernova cosmology cookbook: Bayesian numerical recipes}.
 \emph{arXiv} \textbf{2015}, arXiv:1503.03844.


\bibitem[Mohayaee et al.(2021)]{Mohayaee:2021jzi}
Mohayaee, R.; Rameez, M.; Sarkar, S.\ 
{Do supernovae indicate an accelerating universe?}
{\textit{Eur. Phys. J. ST} \textbf{2021}, \textit{230}, 2067}.
\url{https://doi.org/10.1140/epjs/s11734-021-00199-6}.

 
\end{thebibliography}


\PublishersNote{}
\end{adjustwidth}

\end{document}